\documentclass[letterpaper]{aa}

\usepackage{amssymb,amsmath}
\usepackage{graphicx,epsfig}
\usepackage{subfigure}
\usepackage{natbib}
\usepackage{array}
\bibpunct{(}{)}{;}{a}{}{,} % to follow A&A style

\begin{document}

\title{Type Ia supernova rate at a redshift of $\sim$ 0.1}

\author{
G.~Blanc\inst{1,12,22},
C.~Afonso\inst{1,4,8,23},
C.~Alard\inst{24},
J.N.~Albert\inst{2},
G.~Aldering\inst{15,28},
A.~Amadon\inst{1},
J.~Andersen\inst{6},
R.~Ansari\inst{2},
\'E.~Aubourg\inst{1},
C.~Balland\inst{13,21},
P.~Bareyre\inst{1,4},
J.P.~Beaulieu\inst{5},
X.~Charlot\inst{1},
A.~Conley\inst{15,28},
C.~Coutures\inst{1},
T.~Dahl\'en\inst{19},
F.~Derue\inst{13},
X.~Fan\inst{16},
R.~Ferlet\inst{5},
G.~Folatelli\inst{11}, 
P.~Fouqu\'e\inst{9,10},
G.~Garavini\inst{11},
J.F.~Glicenstein\inst{1},
B.~Goldman\inst{1,4,8,23},
A.~Goobar\inst{11},
A.~Gould\inst{1,7},
D.~Graff\inst{7},
M.~Gros\inst{1},
J.~Haissinski\inst{2},
C.~Hamadache\inst{1},
D.~Hardin\inst{13},
I.M.~Hook\inst{25},
J.~de Kat\inst{1},
S.~Kent\inst{18},
A.~Kim\inst{15},
T.~Lasserre\inst{1},
L.~Le Guillou\inst{1},
\'E.~Lesquoy\inst{1,5},
C.~Loup\inst{5},
C.~Magneville \inst{1},
J.B.~Marquette\inst{5},
\'E.~Maurice\inst{3},
A.~Maury\inst{9},
A.~Milsztajn \inst{1},
M.~Moniez\inst{2},
M.~Mouchet\inst{20,22},
H.~Newberg\inst{17},
S.~Nobili\inst{11},
N.~Palanque-Delabrouille\inst{1},
O.~Perdereau\inst{2},
L.~Pr\'evot\inst{3},
Y.R.~Rahal\inst{2},
N.~Regnault\inst{2,14,15},
J.~Rich\inst{1},
P.~Ruiz-Lapuente\inst{27},
M.~Spiro\inst{1},
P.~Tisserand\inst{1},
A.~Vidal-Madjar\inst{5},
L.~Vigroux\inst{1},
N.A.~Walton\inst{26},
S.~Zylberajch\inst{1}.
}

\institute{
%1
DSM/DAPNIA, CEA/Saclay, 91191 Gif-sur-Yvette Cedex, France
\and
%2
Laboratoire de l'Acc\'{e}l\'{e}rateur Lin\'{e}aire,
IN2P3 CNRS, Universit\'e Paris-Sud, 91405 Orsay Cedex, France
\and
%3
Observatoire de Marseille,
2 pl. Le Verrier, 13248 Marseille Cedex 04, France
\and
%4
Coll\`ege de France, Physique Corpusculaire et Cosmologie, IN2P3 CNRS, 
11 pl. M. Berthelot, 75231 Paris Cedex, France
\and
%5
Institut d'Astrophysique de Paris, INSU CNRS,
98~bis Boulevard Arago, 75014 Paris, France
\and
%6
Astronomical Observatory, Copenhagen University, Juliane Maries Vej
30, 2100 Copenhagen, Denmark
\and
%7
Departments of Astronomy and Physics, Ohio State University, Columbus, 
OH 43210, U.S.A.
\and
%8
Department of Astronomy, New Mexico State University, Las Cruces, NM
88003-8001, U.S.A.  
\and
%9
European Southern Observatory (ESO), Casilla 19001, Santiago 19, Chile
\and
%10
Observatoire Midi-Pyr\'en\'ees, 14 avenue Edouard Belin,
31400 Toulouse, France
\and
%11
Department of Physics, Stockholm University, 
AlbaNova University Center, S-106 91 Stockholm, Sweden
\and
%12
Osservatorio Astronomico di Padova, INAF, vicolo dell'Osservatorio 5 - 
35122 Padova, Italy 
\and
%13
Laboratoire de Physique Nucl\'eaire et de Hautes Energies, 
IN2P3 - CNRS - Universit\'es Paris 6 et Paris 7, 
4 place Jussieu, 75252 Paris Cedex 05, France
\and
%14
Laboratoire Leprince-Ringuet,
LLR/Ecole Polytechnique,
Route de Saclay,
91128 Palaiseau CEDEX,
France
\and
%15
Lawrence Berkeley National Lab,
1 Cyclotron Road,
Berkeley, CA 94720, U.S.A.
\and
%16
Steward Observatory,
The University of Arizona,
933 N. Cherry Ave,
Tucson, AZ 85721-0065,
U.S.A.
\and
%17
Rensselaer Polytechnic Institute,
110 Eighth Street,
Troy, NY 12180, U.S.A.
\and
%18
Fermilab
Wilson and Kirk Roads,
Batavia, IL 60510-0500, U.S.A.
\and
%19
Space Telescope Science Institute, 3700 San Martin Dr., Baltimore, MD
21218, U.S.A.
\and
%20
LUTH, Observatoire de Paris,
5, Place Jules Janssen,
92195 Meudon Cedex, France
\and
%21
Institut d'Astrophysique Spatiale,
B\^atiment 121, Universit\'e Paris 11,
91405 Orsay Cedex, France
\and
%22
Universit\'e Paris 7 Denis Diderot,
2, place Jussieu, 75005 Paris, France
\and
%23
NASA/Ames Research Center, 
Mail Stop 244, 
Moffet Field, CA 94035, U.S.A.
\and
%24
GEPI, Observatoire de Paris, 77 avenue de l'Observatoire,
75014 Paris, France
\and
%25
Department of Physics, University of Oxford, Nuclear and Astrophysics
laboratory, Keble Road, Oxford OX1 3RH, UK 
\and
%26
Institute of Astronomy, Madingley Road, Cambridge CB3 0HA, UK
\and
%27
Department of Astronomy, University of Barcelona, Barcelona, Spain
\and
%28
Visiting astronomer, Cerro Tololo Inter-American
Observatory, National Optical Astronomy Observatories, which are
operated by the Association of Universities for Research in Astronomy,
under contract with the National Science Foundation.
}

\offprints{G. Blanc, \email{blanc@pd.astro.it}}
\date{Received / Accepted}

\abstract{ We present the type~Ia rate measurement based on two EROS
supernova search campaigns (in 1999 and 2000).  Sixteen supernovae
identified as type~Ia were discovered.  The measurement of the
detection efficiency, using a Monte Carlo simulation, provides the
type~Ia supernova explosion rate at a redshift $\sim$ 0.13. The result
is $0.125^{+0.044+0.028}_{-0.034-0.028}\ h_{70}^2$ SNu where 1 SNu = 1
SN / $10^{10}\ \textrm{L}_{\odot}^B$ / century. This value is
compatible with the previous EROS measurement \citep{hardin2000}, done
with a much smaller sample, at a similar redshift.  Comparison with
other values at different redshifts suggests an evolution of the
type~Ia supernova rate.

}

\authorrunning{G. Blanc et al.}
\titlerunning{Type Ia supernova rate at a redshift of $\sim$ 0.1}

\maketitle

\section{Introduction}

Type~Ia supernovae are believed to be thermonuclear explosions of a
white dwarf reaching the Chandrasekhar mass after accreting matter in
a binary system (see \textit{e.g.} \cite{livio2001}). They are the
main mechanism to enrich the interstellar medium (ISM) with iron-peak
elements (with $\sim$ 0.6 M$_{\odot}$ of nickel released per event on
average). The knowledge of the number of such events per galaxy and
per unit of time is crucial to understand the matter cycle and the ISM
chemistry. From a cosmological point of view the measurement of the
evolution of the explosion rate will put strong constraints on galaxy
evolution. Measuring the evolution of the core-collapse supernovae
(type~II and Ib/c) explosion rate would give a new unbiased view on
the star formation rate (SFR) history, independent of any peculiar
tracer, because gravitational supernovae are directly linked to
massive star formation. Up to now, the type~II supernova rate has been
measured only in the local Universe \citep{cappellaro1999}. On the
other hand, the measurement of the evolution of the explosion rate of
type~Ia supernova would give strong constraints on the SFR history and
on many other parameters such as the progenitor evolution (birth rate
of binaries, time delay between the white dwarf formation and the
thermonuclear explosion...) or the nature of the progenitor itself.

This paper is organized as follows: first the EROS search for
supernovae is reviewed in Sect.~\ref{sec:eros}. Then the two supernova
search campaigns at the origin of this work are presented in
Sect.~\ref{sec:campaigns}.  Sections~\ref{sec:rate} to
\ref{sec:integral} deal with the rate computation.  Our results are
presented in Sect.~\ref{sec:result} and discussed in
Sect.~\ref{sec:discussion}.

\section{The EROS search for supernovae}
\label{sec:eros}

The EROS experiment used a 1-meter telescope based at La Silla
observatory, Chile, designed for a baryonic dark matter search using
the microlensing effect \citep[see
\textit{e.g.}~][]{lasserre2000,afonso2003b,afonso2003a}. The telescope
beam light was split by a dichroic cube into two wide field cameras (1
deg$^2$). The cube sets the passband of both cameras, a ``blue'' one
overlapping the $V$ and $R$ standard bands, another ``red'' one
matching roughly the $I$ standard band\footnote{By ``standard'' we
mean the Johnson-Cousins $UBVRI$ photometric system
\citep{johnson1965,cousins1976} as used by \cite{landolt1992}.}. Each
camera was a mosaic of eight 2048 $\times$ 2048 pixels CCDs with a
pixel size of $0.60''$ on the sky
\citep{bauer1997,palanque-delabrouille1998}. In the following, we will
use only the blue camera. One ``image'' stands for one CCD, and one
``field'' is for the entire camera with its eight CCDs, \textit{i.e.}
$\sim$ 1 deg$^2$.

A wide field camera is a good asset to perform supernova searches as
the number $n$ of detected supernovae is proportional to $ \Omega
\cdot z_{\text{lim}}^2$ where $\Omega$ is the solid angle surveyed and
$z_{\text{lim}}$ is the limiting redshift of the search -- typically
EROS detected type~Ia supernovae up to $z = 0.3$. EROS dedicated about
15~\% of its time to supernova search between 1997 and 2000. The main
limitation has been to secure enough telescope time to perform the
photometric and spectroscopic follow-up of discovered events.

To search for supernovae EROS uses the CCD subtraction technique which
consists in observing the same fields three weeks to one month apart
in order to detect transient events and catch only supernovae near
maximum light (a typical type~Ia supernova reaches its maximum light
in the optical wavelength roughly 15 to 20 days after the
explosion). Both observations are done around new moon in order to
limit the sky background. The two sets of images are then
automatically subtracted after a spatial alignment, a flux alignment
and a seeing convolution. 

Cuts are applied to eliminate known classes of variable
objects such as asteroids, variable stars, and cosmic rays.  The most
important cut is the requirement that the transient object be near an
identified galaxy.  More precisely, it is required to be within an
ellipse centered on any galaxy with semi-major and minor axes of
length 8 times the r.m.s.  radius of the galaxy's flux distribution
along each axis. This ellipse should contain essentially all
supernovae associated with a galaxy since $\sim 99$~\% of galactic
light is contained within an ellipse of size 3 times the
r.m.s. radius.   
To eliminate cosmic rays, each exposure consists in
two subexposures: SExtractor \citep{bertin1996} software is run on
both images to make a cosmic-ray catalog before adding the two images.

In spite of these cuts, a human eye is still needed to check the
remaining candidates and discard obvious artefacts (residual cosmic
rays, bad subtractions, asteroids, variable stars...). Ten to fifteen
search fields are observed every night and the human scanner has to
deal with about 100 candidates the following day.  The serious
candidates are observed again on the following night for
confirmation. Then a spectrum is taken to obtain the supernova type,
phase and redshift.

EROS supernova search fields lay near the celestial equator
(declination between $-4.5^{\circ}$ and $-12^{\circ}$) to be reached
from both hemispheres and between 10$^{\text{h}}$ and
14.5$^{\text{h}}$ in right ascension. They overlap some of the Las
Campanas Redshift Survey (LCRS) \citep{shectman1996} fields which will
facilitate the galaxy sample calibration.

\section{The 1999 and 2000 campaigns}
\label{sec:campaigns}

\begin{table*}[!htb]
\centering
\begin{tabular}{lcccclccc}
\hline
\hline
\textbf{IAU Name}& \textbf{Date (UT)} & $\boldsymbol{V}$ &
$\boldsymbol{\alpha_{J2000.0}}$ & $\boldsymbol{\delta_{J2000.0}}$ & 
\multicolumn{1}{c}{\textbf{z}} 
& \multicolumn{1}{c}{\textbf{Type}} & \multicolumn{1}{c}{\textbf{Phase 
    (days)}}  & \textbf{IAUC}\\
\hline
SN~1999ae & 1999-02-10 & 20.7  & 11 51 24.48 & -04 39 09.4  &   0.076 &
II &     $<+30$  &     7117  \\ 
SN~1999af & 1999-02-12 & 19.2  & 13 44 50.95 & -06 40 12.6  &   0.097 &
Ia &    $-10$    &     7117  \\ 
SN~1999ag & 1999-02-12 & 20.1  & 12 15 22.81 & -05 18 12.4  &   0.099 &
II &           &     7117  \\ 
SN~1999ah & 1999-02-13 & 20.5  & 12 09 37.20 & -06 18 34.3  &   0.080 &
SN? & &     7117-18  \\ 
SN~1999ai & 1999-02-15 & 18.0  & 13 14 10.57 & -05 35 43.7  &   0.018 &
II & $\sim$ +14  &     7117-18 \\ 
SN~1999aj & 1999-02-17 & 20.9  & 11 22 39.34 & -11 43 53.9  & 0.186   &
Ia    &  &                  7117  \\ 
SN~1999ak & 1999-02-17 & 18.5  & 11 06 52.05 & -11 39 13.3  &   0.055 &
Ia & $\sim$ +14  &     7117-18  \\ 
SN~1999al & 1999-02-21 & 19.2  & 11 10 25.68 & -07 26 37.0  &   0.065 &
Ic &    -9     &     7117-18  \\ 
\hline
SN~1999bi & 1999-03-10 & 19.8  & 11 01 15.76 & -11 45 15.2  &   0.124 &
Ia &     +2     &     7136  \\ 
SN~1999bj & 1999-03-10 & 20.5  & 11 51 38.39 & -12 29 08.3  &   0.16  &
Ia &    +17    &     7136  \\ 
SN~1999bk & 1999-03-14 & 18.9  & 11 28 52.01 & -12 18 08.3  &   0.096 &
Ia &    +1     &     7136  \\ 
SN~1999bl & 1999-03-14 & 20.7  & 11 12 13.60 & -05 04 44.8  &   0.300 &
Ia &     0     &     7136  \\ 
SN~1999bm & 1999-03-17 & 20.0  & 12 45 00.84 & -06 27 30.2  &   0.143 &
Ia &     -1     &     7136  \\ 
SN~1999bn & 1999-03-16 & 19.6  & 11 57 00.40 & -11 26 38.4  &   0.129 &
Ia &     -5     &     7136  \\ 
SN~1999bo & 1999-03-17 & 19.5  & 12 41 07.48 & -05 57 25.8  &   0.130 &
Ia &    $>0$     &     7136  \\ 
SN~1999bp & 1999-03-19 & 18.6  & 11 39 46.42 & -08 51 34.8  &   0.077 &
Ia &    -4     &     7136  \\ 
SN~1999bq & 1999-03-19 & 20.7  & 13 06 54.46 & -12 37 11.6  &   0.149 &
Ia &     -1     &     7136  \\ 
\hline
\hline
SN~2000bt & 2000-03-26 &  19.4 & 10 16 18.05 & -05 44 47.3  &   0.04  &
Ia & +20          &  7406  \\
SN~2000bu & 2000-03-31 &  19.4 & 11 27 11.45 & -06 23 14.6  &   0.05  &
II? &  $\sim$ 0    &  7406  \\
SN~2000bv & 2000-04-01 & 20.6  & 12 59 28.70 & -12 20 07.6  &   0.12  &
II? &  $\sim$ 0    &  7406  \\ 
SN~2000bw & 2000-04-04 & 20.5  & 11 09 49.85 & -04 24 46.4  &   0.12  &
II? &  $\sim$ 0    &  7406  \\ 
SN~2000bx & 2000-04-06 &  19.2 & 13 48 55.55 & -06 18 35.9  &   0.09  &
Ia &  $\sim$ 0    &  7406  \\
SN~2000by & 2000-04-07 & 19.2  & 11 39 54.91 & -04 22 16.4  &   0.10  &
Ia &  $\sim$ 0    &  7406  \\
SN~2000bz & 2000-04-08 & 21.2  & 14 15 02.66 & -06 17 16.0  &   0.26  &
Ia &  $\sim$ 0    &  7406  \\
\end{tabular}
\caption{List of supernovae discovered by EROS in 1999 and 2000. From
  left to right: IAU name of each event, discovery date, discovery $V$
  magnitude, SNe celestial coordinates, redshift, type, phase
  (discovery time from maximum light) and IAU.}
\label{tab:sneros}
\end{table*}
At the beginning of 1999, the \textit{Supernova Cosmology Project}
(SCP) undertook a large nearby supernova search involving eight
supernova search groups, including EROS. The goal was to gather a
large set of well sampled, CCD discovered, nearby ($z \lesssim 0.15$)
type~Ia supernovae in order to calibrate the distant events used to
measure the cosmological parameters. The whole collaboration obtained
37 SNe out of which 19 were type~Ia SNe near or before maximum light
and were followed both spectroscopically and photometrically.  EROS
observed 428 square degrees of sky in two steps and discovered 12
type~Ia (cf. IAUC 7117 and 7136) of which 7 have been followed by the
collaboration. After obtaining reference images of search fields in
January 1999 (01/12 to 01/29), the SN search was conducted between
February 4 and February 27 (new moon was on February 16), and between
March 09 and March 27 (new moon was on March 17).

One year later, EROS was involved with the \textit{European Supernova
Cosmology Consortium} (ESCC, involving French, British, Swedish and
Spanish institutes) to search for intermediate redshift ($z \sim 0.2 -
0.4$) supernovae.  Four type~Ia supernovae were discovered (cf. IAUC
7406). One hundred and seventy square degrees were observed between
2000, March 27 and April 9 (new moon on April 4); reference images
were taken between February 27 and March 15.  The main difference
between the two searches was the exposure time used: for the 1999
campaign it was set to 300 seconds, while for the 2000 search it was
600 seconds allowing a deeper search by 19~\% in redshift (as
$z_{\text{lim}} \propto T_{\text{exp}}^{1/4}$, where $T_{\text{exp}}$
the exposure time).

In both campaigns, spectra of each candidate were taken. The main
characteristics  of the discovered supernovae are summarized in
Table~\ref{tab:sneros}.

\section{Principle of the measurement}
\label{sec:rate}

Our supernovae search requires that a supernova be associated with an
identified host galaxy.  We must therefore make an assumption about
how the supernova rate scales with galaxy luminosity, so as to correct
the rate for supernovae in dim, unidentified galaxies.  We choose to
assume that the number of supernovae per galxay and per unit time is
proportional to the red galaxy luminosity, an assumption that receives
some empirical support, though mostly in the blue band
\citep{tammann1970,cappellaro1993}. Under this assumption, the
explosion rate $\mathcal{R}_{\text{SN}}$ is given by the ratio of the
number of detected supernovae of type~Ia $\mathcal{N}_{\text{SNe}}$,
to the number of galaxies to which the search is sensitive
$\mathcal{N}_{\text{gal}}$, weighted by their mean luminosity $\langle
\mathcal{L}_{\text{gal}}\rangle$ and by the mean time interval
$\langle \mathcal{T}\rangle$ during which the supernovae can be
detected:
\begin{equation}
\mathcal{R}_{\text{SN}} =
  \frac{\mathcal{N}_{\text{SNe}}}{\mathcal{N}_{\text{gal}} \cdot
  \langle \mathcal{L}_{\text{gal}}\rangle
  \cdot \langle \mathcal{T}\rangle}.
\label{eqn:basetaux}
\end{equation}
The time interval or \textit{control time} $\mathcal{T}$ during which
a given supernova is visible depends on the supernova detection
efficiency $\varepsilon$.  Since the efficiency itself depends on many
parameters, the observed control time $\mathcal{T}$ for a supernova of
redshift $z$ is given by an integral of $\varepsilon$ over these
parameters:
\begin{equation}
\mathcal{T}(z) = \int_{\phi_{\text{SNIa}}}
\int^{+\infty}_{-\infty} \varepsilon(\Delta ; m(t,z;
\phi_{\text{SNIa}}))\ dt\ d\phi_{\text{SNIa}},
\label{eqn:tc}
\end{equation} 
where $\Delta$ is the time interval between the search images and the
reference images; $m(t,z; \phi_{\text{SNIa}})$ is the supernova
apparent magnitude ($t$ is the time from maximum light or phase; $z$
is the redshift). The magnitude $m$ depends also indirectly on the
light curve distribution of type~Ia supernova (see \S~\ref{sec:snlf}),
that we note $\phi_{\text{SNIa}}$.  As we know the galaxy sample in
the search images (see \S~\ref{sec:galaxies}), but not their
redshifts, the restframe explosion rate can be rewritten as:
\begin{equation}
\mathcal{R}_{\text{SN}} = \frac{\mathcal{N}_{\text{SNe}}}{\sum_g 
\int_0^{+\infty} p(z|m_g)\cdot
\mathcal{L}_g(m_g,z) \cdot
\mathcal{T}(z)/(1+z) \cdot dz},
\label{eqn:tauxgene}
\end{equation}
where $g$ is a subscript running on all galaxies of all the search
images; $p(z|m_g)$ is the probability density of the redshift $z$ of a
galaxy knowing its apparent magnitude $m_g$ (see \S~\ref{sec:pzm});
$\mathcal{L}_g$ is the galaxy luminosity (which depends on its
redshift and apparent magnitude, as well as on a cosmological model:
we use a standard model, ($\Omega_{M_{\circ}},\,
\Omega_{\Lambda_{\circ}}$)~=~(0.3,~0.7); for the Hubble constant, we
use the notation $H_{\circ}~=~70\ h_{70}\ \text{km}\ \text{s}^{-1}\
\text{Mpc}^{-1}$ wherever needed).  Different steps are needed to
calculate the denominator of eq.~(\ref{eqn:tauxgene}). The first one
is to establish a well-defined sample of galaxies in the search
images. This is then used to measure the detection efficiency. Each of
these steps is detailed in the following sections.

In the following the EROS instrumental magnitude will be denoted:
\begin{equation}
\widetilde{\mathcal{B}}_{\text{eros}} \equiv -2.5\cdot \log
   \frac{\mathcal{F}_{ADU}}{T_{\text{exp}}}
\label{eqn:berosnot}
\end{equation}
where $\mathcal{F}_{ADU}$ is the flux as measured in CCD images in
digital units; $T_{\text{exp}}$ is the exposure time in
seconds. Throughout this paper the magnitudes in EROS images are
computed using an adaptive aperture photometry as provided by the
SExtractor software \citep{bertin1996}, which is accurate enough for
our purpose.  Photometry for stellar objects was done using
observations of Landolt stars, as detailed in
Sect. \ref{sec:calibration}.  Galactic magnitudes needed to derive the
redshift distribution and to normalize the SN rate was calibrated
using LCRS $R$ magnitudes.

\section{The galaxy sample}
\label{sec:galaxies}

\begin{figure}[!htb]
\centering
\includegraphics[width=8.5cm]{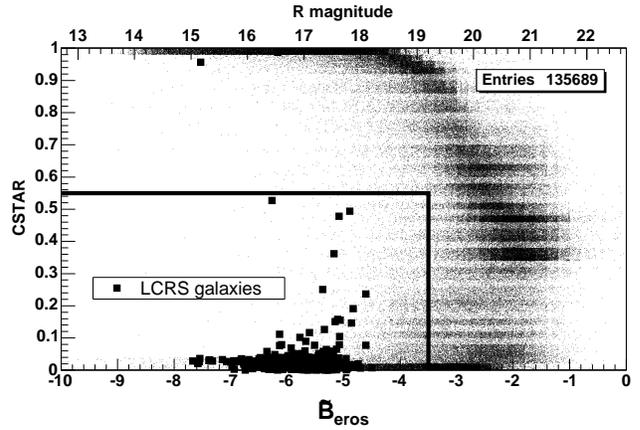}
\caption{EROS magnitude distribution vs. value of CSTAR. Objects
  with CSTAR around 0.5 are faint and thus difficult to
  classify. Squares stand for LCRS galaxies. The dark line represents
  the cut (\ref{eqn:cutgal99}) for the 1999 search.}
\label{fig:mag-cstar1999}
\end{figure}

We use the SExtractor software \citep{bertin1996} to detect objects on
the images.  Objects are classified as stars and galaxies using the
SExtractor CSTAR parameter which quantify the ``stellarness'' of an
object. For stars CSTAR $\rightarrow$ 1, and for galaxies CSTAR
$\rightarrow$ 0. The critical input in order to obtain a reliable
value of this parameter is the image seeing.  SExtractor is then run
twice on the images, first to obtain the seeing, and a second time to
compute CSTAR. Figure~\ref{fig:mag-cstar1999} shows a scatter plot of
CSTAR versus the galaxy magnitude. We see that the galaxy star
separation is clear for $R < 19$. The galaxies of the LCRS catalog
\citep{shectman1996} shown in the plot allow one to verify the
efficiency of the classification for $R < 18$.
\begin{figure*}[!htb]
\centering
\subfigure[Galaxy count on EROS fields done on 180 images from the 1999
  search. The completeness limit is for CSTAR $<$ 0.55 and
$\widetilde{\mathcal{B}}_{\text{eros}} < -2.6$.]{  
% FICHIER $SN_NT_DIR/cur/nr1999.nt
% commande root : .x comptage.C
\includegraphics[width=8cm]{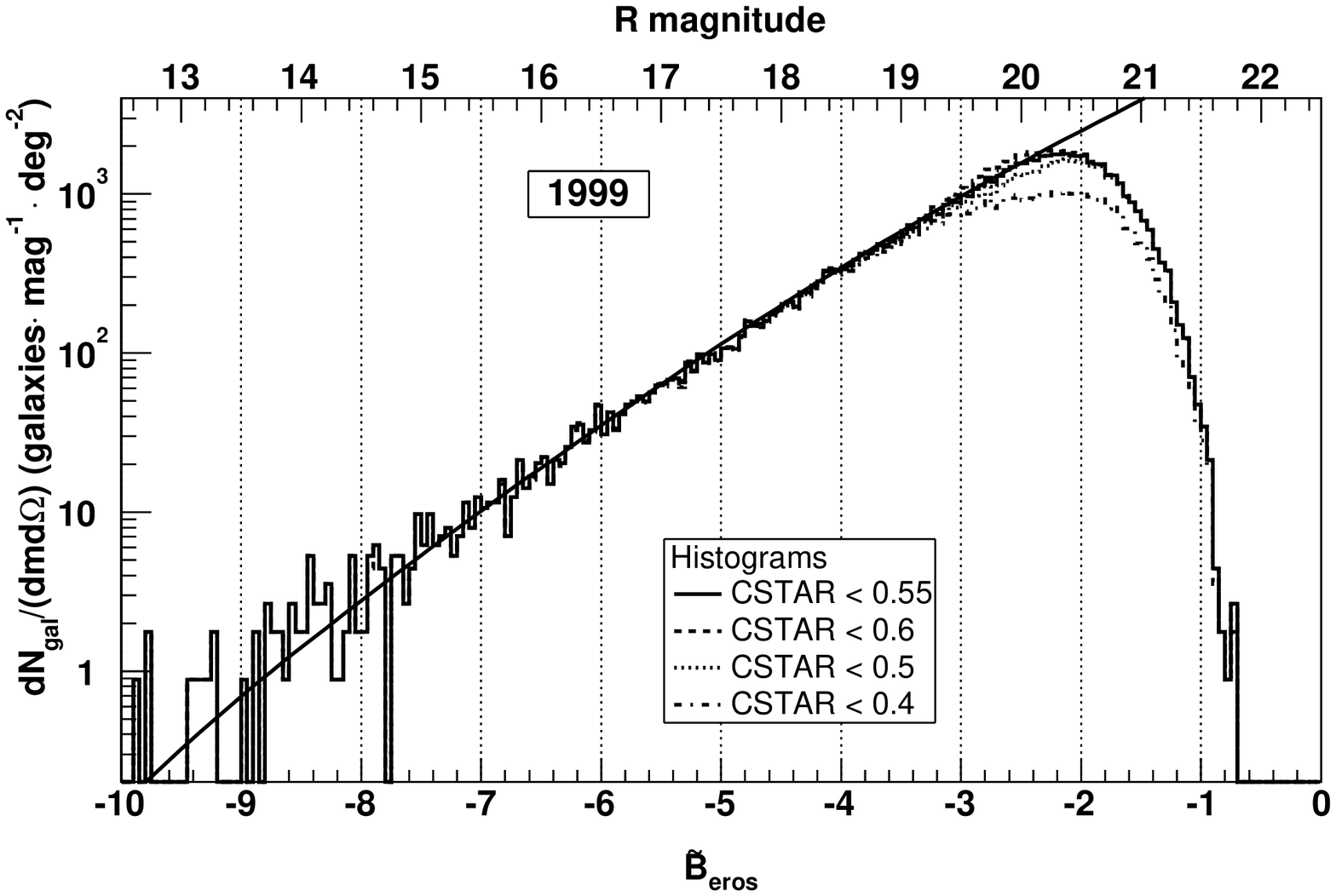}
\label{fig:comptegal1999a}}
% commande root : .x comptage_zoom.C
\subfigure[Galaxy count on EROS fields done on 29 images from the 2000
  search. The completeness limit is for CSTAR $<$ 0.55 and
$\widetilde{\mathcal{B}}_{\text{eros}} < -2.2$.]{
% FICHIER $SN_NT_DIR/cur/nr1999.nt
% commande root : .x comptage.C
\includegraphics[width=8cm]{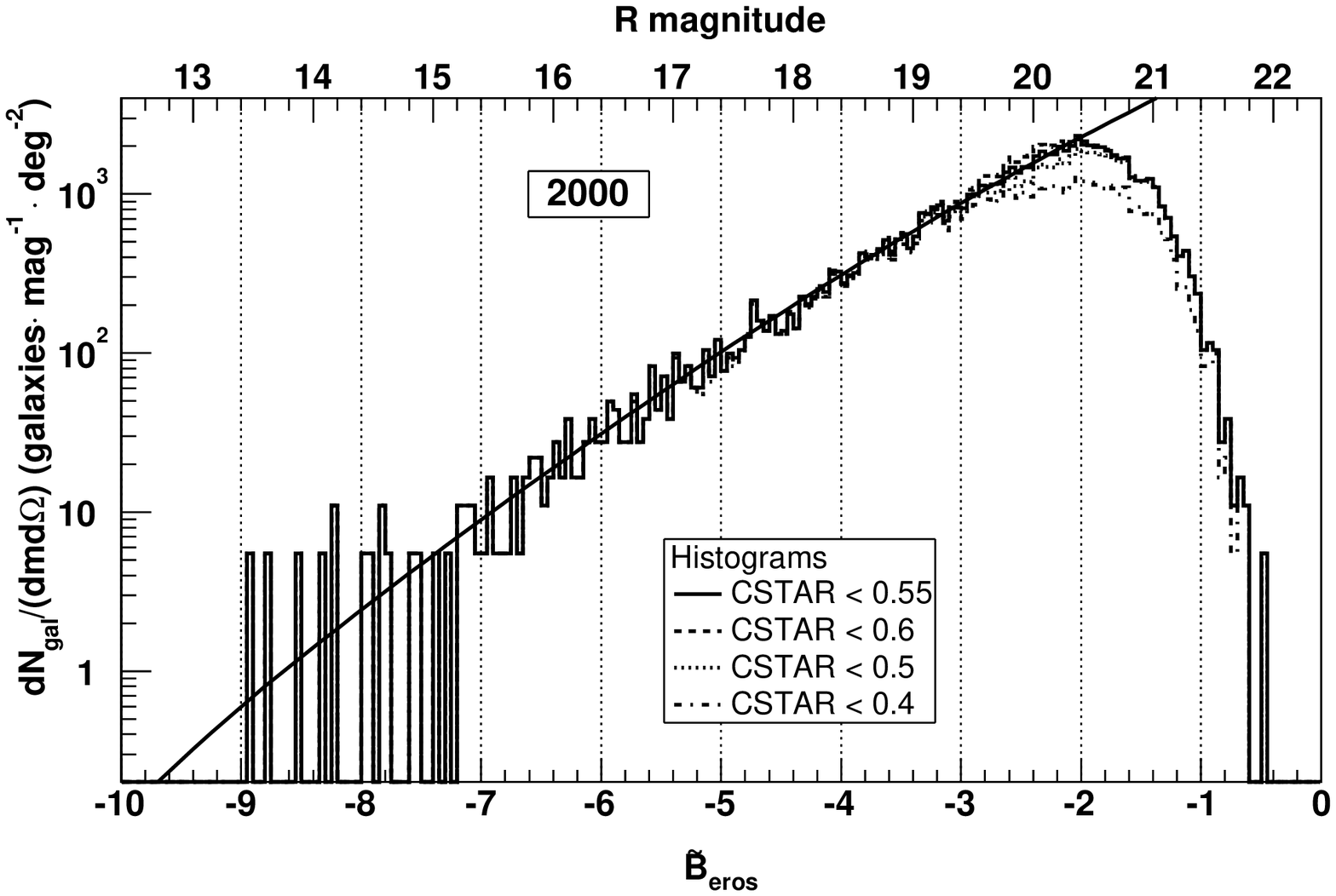}
\label{fig:comptegal2000a}}
\caption{Galaxy counts in EROS supernova fields for various cuts on
the CSTAR parameter. The nominal value for this cut is CSTAR $<$
0.55. Beyond, faint stars pollute the galaxy sample. Below, faint
galaxies are missed. The curve stands for published data such as the
count in $R$ band by \cite{bertin1997}. The matching of the curve with
the histogram confirms our galaxy magnitude calibration,
eq.~(\ref{eqn:zpgal}). The main difference between the 1999 and the
2000 searches was the exposure time, set to 300 seconds in 1999 and 600
seconds in 2000.}
\label{fig:comptegal}
\end{figure*}

To further check the reliability of the galaxy sample we make a
``galaxy count'' that we compare to published ones. As illustrated in
Fig.~\ref{fig:comptegal} we are able to define accurately the
completeness limit of our survey as the point where our data turn down
relative to the published counts.  At this point, stars and galaxies
are quite hard to distinguish: most of the objects still have a CSTAR
$\sim$ 0.5 (see Fig.~\ref{fig:mag-cstar1999} for
$\widetilde{\mathcal{B}}_{\text{eros}} \sim -2.6$). To take this into
account, we set the galaxy magnitude cut one magnitude below the
completeness limit in order to prevent contamination of our galaxy
sample by stars.  Then:
\begin{equation}
\text{for}\ 1999\ 
\left\{
\begin{array}{l}
\text{\index{CSTAR}CSTAR} < 0.55, \\
\widetilde{\mathcal{B}}_{\text{eros}} < -3.5;\ (\textit{i.e.}\ R <
19),  \\ 
\end{array}
\right.
\label{eqn:cutgal99}
\end{equation}
\begin{equation}
\text{for}\ 2000\ 
\left\{
\begin{array}{l}
\text{\index{CSTAR}CSTAR} < 0.55, \\
\widetilde{\mathcal{B}}_{\text{eros}} < -3.1;\ (\textit{i.e.}\ R <
19.3). \\ 
\end{array}
\right.
\label{eqn:cutgal00}
\end{equation}
The effect of this cut is shown in Fig.~\ref{fig:mag-cstar1999}. We
used the LCRS galaxy catalog \citep{shectman1996} to check it, as
illustrated in Fig.~\ref{fig:mag-cstar1999}.  This cut removes events
in faint galaxies which are visually indistinguishable from variable
stars.

According to the LCRS luminosity function \citep{lin1996}, our cut
requiring $R<19$ means that at our mean redshift of 0.13 (see
Fig.~\ref{fig:distzobs}), our supernova search is sensitive to
supernovae in galaxies that generate about 85~\% of the total stellar
luminosity.  We are not sensitive to supernovae in the multitude of
low-luminosity galaxies but these galaxies generate little total
light. If the supernova rate is proportional to the luminosity, they
also generate few supernovae.

\section{The detection efficiency}
\label{sec:efficiency}

\begin{figure}[!htb]
\centering
\includegraphics[width=8cm]{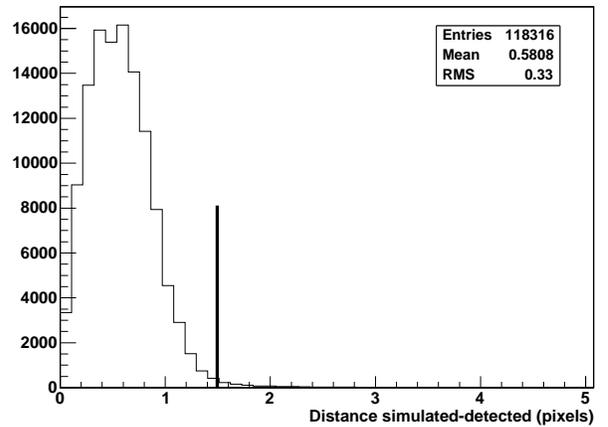}
\caption{Histogram of the distance between the simulated supernovae
  and the detected ones. The cut we used to identify the detected
  population is set at a radius of 1.5 pixels from the simulated
  supernova as shown by the dark line.}
\label{fig:damc1999}
\end{figure}

\begin{figure}[!htb]
\centering
\includegraphics[width=8.5cm]{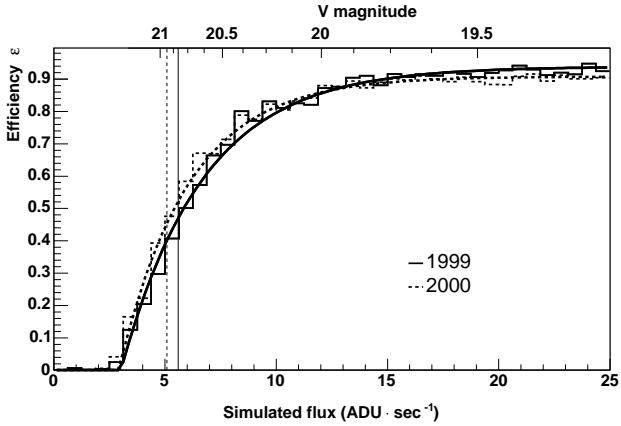}
\caption{Mean efficiency computed for each of the two
searches as a function of the simulated flux, in ADU s$^{-1}$ or $V$
magnitude (according to eq.~(\ref{eqn:zpsne})), for both the 1999
search (continuous line) and the one of 2000 (dashed line). The curves
are the fits.  With its exposure time twice the one of 1999, the 2000
search went about 0.2 mag deeper. The two vertical lines represent the
limiting magnitudes defined as the magnitude at half the maximum
efficiency: 5.59 ADU s$^{-1}$ for 1999 and 5.09 ADU s$^{-1}$ for
2000. The ratio between the two limits is 1.10, to be compared with
1.41 expected for an exposure time ratio of 2. The plateaux reached at
high flux are respectively 0.94 and 0.91 for 1999 and 2000.}
\label{fig:efficiency}
\end{figure}

\begin{figure}[!htb]
\centering
\includegraphics[width=8.5cm]{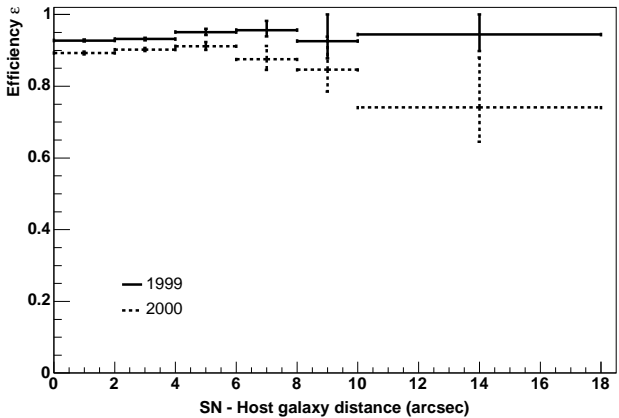}
\caption{Mean efficiency as a function of the distance of simulated
supernovae to their host galaxy peaks, computed for each of the two
searches. Since there is a loss of efficiency at low fluxes, this plot
is done for fluxes above 20 ADU s$^{-1}$ for 1999 and above 13 ADU
s$^{-1}$ for 2000.}
\label{fig:effdist}
\end{figure}

In order to measure the detection efficiency, we superimpose simulated
supernovae on the galaxies.  On each search image we choose a bright
-- but not saturated -- star (with a signal to noise ratio typically
above 50) as a template to be ``copied'' onto the potential host
galaxy.  Among all such stars found in the image, we take the one
nearest to the galaxy on which we wish to add a supernova. We move
only the pixels included in a circle with a radius given by the point
where the star flux calculated by using a gaussian is equal to half a
standard deviation of the sky background. Each pixel is rescaled to
the desired flux (simulated supernova flux is distributed uniformly
between 0 and 15\,000 ADU -- the latter corresponding roughly to $V
\sim 18.5$ for 1999 and to $V \sim 19.2$ for 2000).  Simulated
supernovae are placed at random positions within the isophotal limits
of the galaxy, \textit{i.e.} inside an ellipse centered on the galaxy
and that has semi-major and minor axes of length 3 times the r.m.s. of
the galaxy's flux along the axis distribution; this ellipse contains
$\sim$99~\% of the galaxy flux. The positions are chosen following a
probability proportional to the local surface brightness.
They are placed on the galaxies of the images before the search
software is run. A simulated supernova is detected by the search
software if its signal-to-noise ratio is greater than 5 and if it is
situated within a radius of 1.5 pixels of the simulated one, according
to Fig.~\ref{fig:damc1999}.  The resulting efficiency as a function of
the simulated SN flux is the ratio between detected supernovae and
simulated ones. The result is fitted by an analytic function in order
to smooth the histogram.
The efficiency used in eq.~(\ref{eqn:tc}) is computed for each search
image. A ``global'' efficiency computed for each of the two campaigns
as a function of simulated flux is shown in Fig.~\ref{fig:efficiency}
with the resulting fits. Figure~\ref{fig:effdist} shows this
``global'' efficiency as a function the simulated SN distances to
their hosts.  We remark that we observe no significant efficiency loss
in the center of galaxies. This improvement comes from using CCD
detectors instead of photographic plates (where the core of galaxies
saturates -- see {\it e.g.}  \cite{howell2000}). Moreover the
efficiency is roughly independent of the position within the central
$\sim 8''$ of the galaxies.
Beyond this radius, corresponding to the isophotal limit of most
galaxies in our sample, few supernovae are simulated leading to the
large statistical fluctuations in Fig.~\ref{fig:effdist}.  Near the
center of galaxies, the efficiency is about 0.9, limited mostly by
``geometrical'' losses near CCD edges and around dead pixels.

\section{The integral computation}
\label{sec:integral}

\subsection{The algorithm}

To compute the integral in the denominator of
eq.~(\ref{eqn:tauxgene}), we have to integrate the detection
efficiency over various parameters, such as the type~Ia supernova
light curve distribution, the supernova redshift and its phase at the
time of discovery (time from maximum light). In order to do that, we
use a Monte-Carlo method: a light curve is randomly drawn among a
selected set of SNe (see Sect.~\ref{sec:snlf}); a redshift is drawn
according to the probability density $p(z|m_g)$ as defined in
Sect.~\ref{sec:pzm}; the selected light curve is adjusted according to
this redshift. A phase is drawn uniformly, which provides a standard
magnitude. After transformation of standard $V$ and $R$ magnitudes
into ADU (see Sect.~\ref{sec:calibration}), the corresponding
efficiency is read from Fig.~\ref{fig:efficiency}.

\begin{figure}[!htb]
\centering
\includegraphics[width=8.5cm]{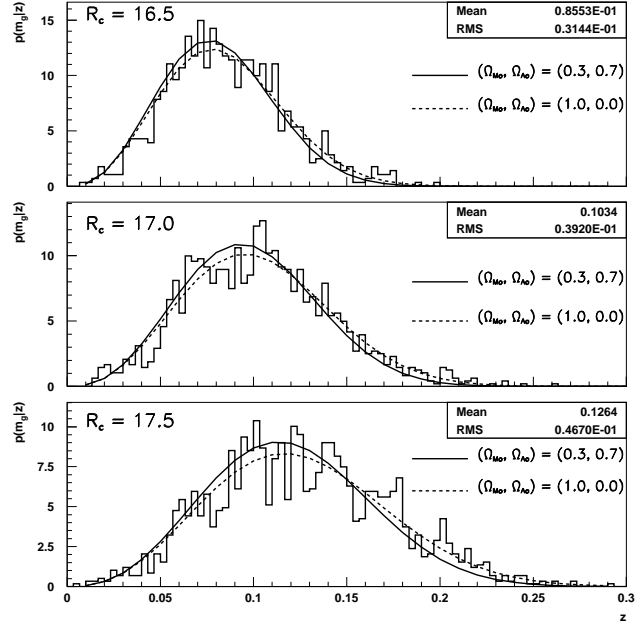}
\caption{Examples of redshift distributions of galaxies knowing their
  apparent magnitude, as given by (\ref{eqn:pmz}).  The solid lines
  show the predicted distribution for three apparent magnitudes $m_g$
  and two sets of cosmological parameters.  The histograms show the
  distribution of LCRS redshifts for galaxies within 0.1 mag of the
  given magnitude.}
\label{fig:distzlcrs}
\end{figure}

\subsection{Redshift distribution of galaxies}
\label{sec:pzm}

We do not know the redshifts of all the galaxies in the images.  But
all we need is the probability distribution $p$ for the redshift $z$
given the galaxy's apparent magnitude $m_g$:
\begin{equation}
p(z|m_g) \varpropto \frac{dN}{dm_gdzd\Omega} =
\frac{dV_c}{d\Omega dz} \cdot \phi_g(M_g)
\label{eqn:pmz}
\end{equation}
where $N$ is the number of galaxies having this redshift and apparent
magnitude, $\Omega$ is the solid angle surveyed, $V_c$ is the comoving
volume and $\phi_g$ the galaxy luminosity function ($M_g$ being their
absolute magnitude). To compute this, we use the luminosity function
of the LCRS \citep{lin1996}.  Figure~\ref{fig:distzlcrs} gives
examples of such a distribution.

\begin{figure}[!htb]
\centering
\includegraphics[width=8.5cm]{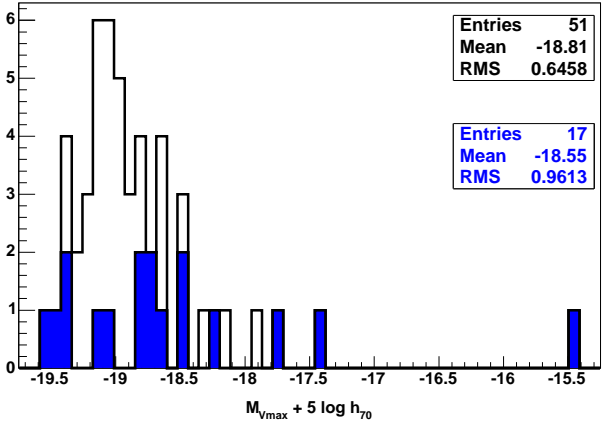}
\caption{Absolute $V$ magnitude at maximum light distribution for the
whole published set (51 SNe, white histogram, upper box) and the
reference sample of 17 supernovae we selected (blue histogram, lower
box). If we remove the lowest luminosity event (SN~1996ai) from this
reference sample, its mean is shifted from $-$18.55 to $-$18.75.}
\label{fig:snelf}
\end{figure}
%dm15nearby-pap.eps
%
\begin{figure}[!htb]
\centering
\includegraphics[width=8.5cm]{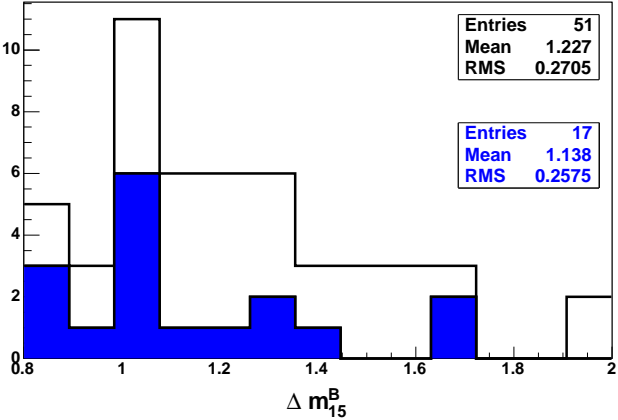}
\caption{Distribution of the light curve shape parameter $\Delta
m_{15}^B$ for the whole published set (51 SNe, white histogram, upper
box) and the reference sample of 17 supernovae we selected (blue
histogram, lower box).}
\label{fig:snelfdm15}
\end{figure}

\subsection{Type Ia supernova light curve distribution}
\label{sec:snlf}

We need to integrate the efficiency on a large and representative
sample of supernova light curves, so as to reproduce the variety of
such objects. The light curve distribution of type~Ia supernova can be
parametrized, at least at first order, using two parameters: the
luminosity at maximum light, and a light curve shape parameter. 
The light curve shape can be quantified using the $\Delta m_{15}^B$
parameter \citep{phillips1993}, that is the magnitude difference
between the light at maximum and 15 days later on the $B$ light curve.

Out of 51 published type~Ia supernovae observed light curves
\citep{hamuy1996c, riess1999}, we selected a reference sample of
seventeen objects that have a good sampling in both $V$ and $R$ bands,
the standard bands used to calibrate the EROS band. Each light curve
is then fitted with an analytic template \citep{contardo2000}, which
allows us to reach any point of the light curve without interpolation;
the selected objects of the reference sample have a very good fit.
The resulting distribution of absolute magnitude at maximum light (the
luminosity function) is shown in Fig.~\ref{fig:snelf}.  The reference
sample we have used has a lower mean luminosity than the full
sample. This is mainly due to one very subluminous event, SN~1996ai
(which may be very extinguished by its host galaxy -- $A_V \sim 4$
according to \cite{tonry2003}).  The $\Delta m_{15}^B$ distribution of
the reference sample is shown in Fig.~\ref{fig:snelfdm15}; it is
similar to that of the full sample.  Our supernova rate will be
presented in a form such that it can be revised if future measurements
give different values for the mean luminosities and $\Delta m_{15}^B$.

\subsection{From standard magnitudes to ADUs}
\label{sec:calibration}

As the detection efficiency is measured as a function of supernova
flux in the EROS band, and as the published light curve sample we will
use to reproduce the supernova light curve distribution is expressed
in standard magnitudes, we need to translate one system into the other
using a calibration relation such as:
\begin{equation}
\widetilde{\mathcal{B}}_{\text{eros}} = V + \alpha \cdot
(V - R) + \beta + \delta \cdot (\xi - 1),
\end{equation}
between the measured flux on the CCD images
($\widetilde{\mathcal{B}}_{\text{eros}}$ is defined by
eq.~(\ref{eqn:berosnot})) and the observed magnitudes through the
standard bands: we choose $V$ and $R$ which are the closest to the
EROS band; $\alpha$ is the color term; $\beta$ is the so-called zero
point; $\delta \cdot (\xi -1)$ ($\xi$ is the airmass) is the
atmospheric absorption: the atmospheric absorption at meridian
(\textit{i.e.} at $\sim 20^{\circ}$ from the zenith) is included in
the zero point.  $V$ and $R$ are related to the intrinsic magnitudes
of the supernova $V_{sn}$ and $R_{sn}$ after correcting for the
various extinctions:
\begin{equation}
V = V_{sn} + A_V + A_V^{\text{extra}}, 
\end{equation}
\begin{equation}
R = R_{sn} + A_R + A_R^{\text{extra}}, 
\end{equation}
where $A_V$ and $A_R$ are the Galactic absorption;
$A_V^{\text{extra}}$ and $A_R^{\text{extra}}$ are the extragalactic
absorption which could be either an intergalactic extinction or a host
galaxy extinction or both.

Moreover, we use light curves of observed nearby supernovae. These
light curves are transformed to any redshift; the relationship between
the published light curve $m_F(t_1,z_1)$ in a filter $F$ and the
``new'' redshifted light curve $m_F(t_2,z_2)$\footnote{The supernova
phase $t_1$ at $z_1$ is related to $t_2$ as $t_1/(1+z_1) =
t_2/(1+z_2)$} is given by
\begin{equation}
\begin{split}
m_F(t_2,z_2) - & m_F(t_1,z_1) = 5 \log
\frac{\mathcal{D}(z_2)}{\mathcal{D}(z_1)} \\
&\quad + K_F(t_2,z_2) - K_F(t_1,z_1),\\
\label{eqn:m1m2}
\end{split}
\end{equation}
where $\mathcal{D}$ is the luminosity distance and $K$ is a term which
takes into account the fact that we observe objects whose flux is
sliding according to their redshift through a fixed passband in
wavelength space. This term is known as the ``$K$-correction'' (see
Sect.~\ref{sec:kcorrection}).  Note that this relation does not depend
on the Hubble constant.  The following sub-sections deal with the
different terms described above.

\subsubsection{Can we calibrate supernovae using stars?}

\begin{figure}[htb]
\centering
\includegraphics[width=8.5cm]{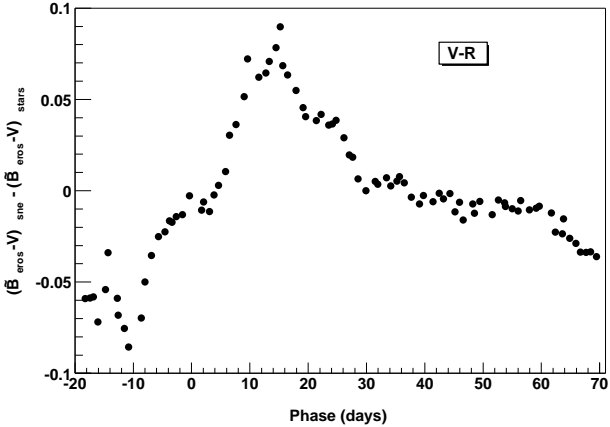}
\caption{Evolution with the supernova phase of the error we made when
  using stars to calibrate supernovae. What is actually plotted is the
  difference between zero points computed for SNe Ia
  $(\widetilde{\mathcal{B}}_{\text{eros}} - V)_{\text{sne}}$ and for
  stars $(\widetilde{\mathcal{B}}_{\text{eros}} - V)_{\text{stars}}$.
  Each dot stands for a star whose color $(V-R)$ is equal to the
  $(V-R)$ color of a SNIa at that phase.  The signal difference is
  never larger than 8~\%. The colors are computed from template
  spectra.}
\label{fig:caltheophaseV}
\end{figure}
As SN spectra are very different from star spectra, showing broad
blended features evolving along the SN phase, we may wonder to what
extent we can use stars to calibrate SN fluxes. In order to quantify
this effect, we use synthetic colors computed with template spectra
for each phase of the SN.  Figure~\ref{fig:caltheophaseV} shows the
error we make when calibrating type~Ia supernova fluxes using stars,
as a function of the supernova phase.  Assuming the color is the same
for SNe and stars ($\alpha_{\text{SNe}} \approx
\alpha_{\text{stars}}$), the plotted quantity
$(\widetilde{\mathcal{B}}_{\text{eros}} - V)_{\text{sne}} -
(\widetilde{\mathcal{B}}_{\text{eros}} - V)_{\text{stars}}$ reflects
directly the zero point differences,
$\beta_{\text{SNe}}-\beta_{\text{stars}}$.  The error is at most 8~\%
around $-$10 days and between +10 and +20 days. We note that for
standard filters, there would be no such error.

\subsubsection{Zero point and color term for the stars}

\begin{figure*}[hbt]
\begin{center}
\subfigure[Zoom on the 1999 supernova search: the mean zero point is
  about 22.7 during the supernova search (see first box for the
  February campaign, and second box for the March campaign).]{
\label{fig:PZ99}
\includegraphics[width=8cm]{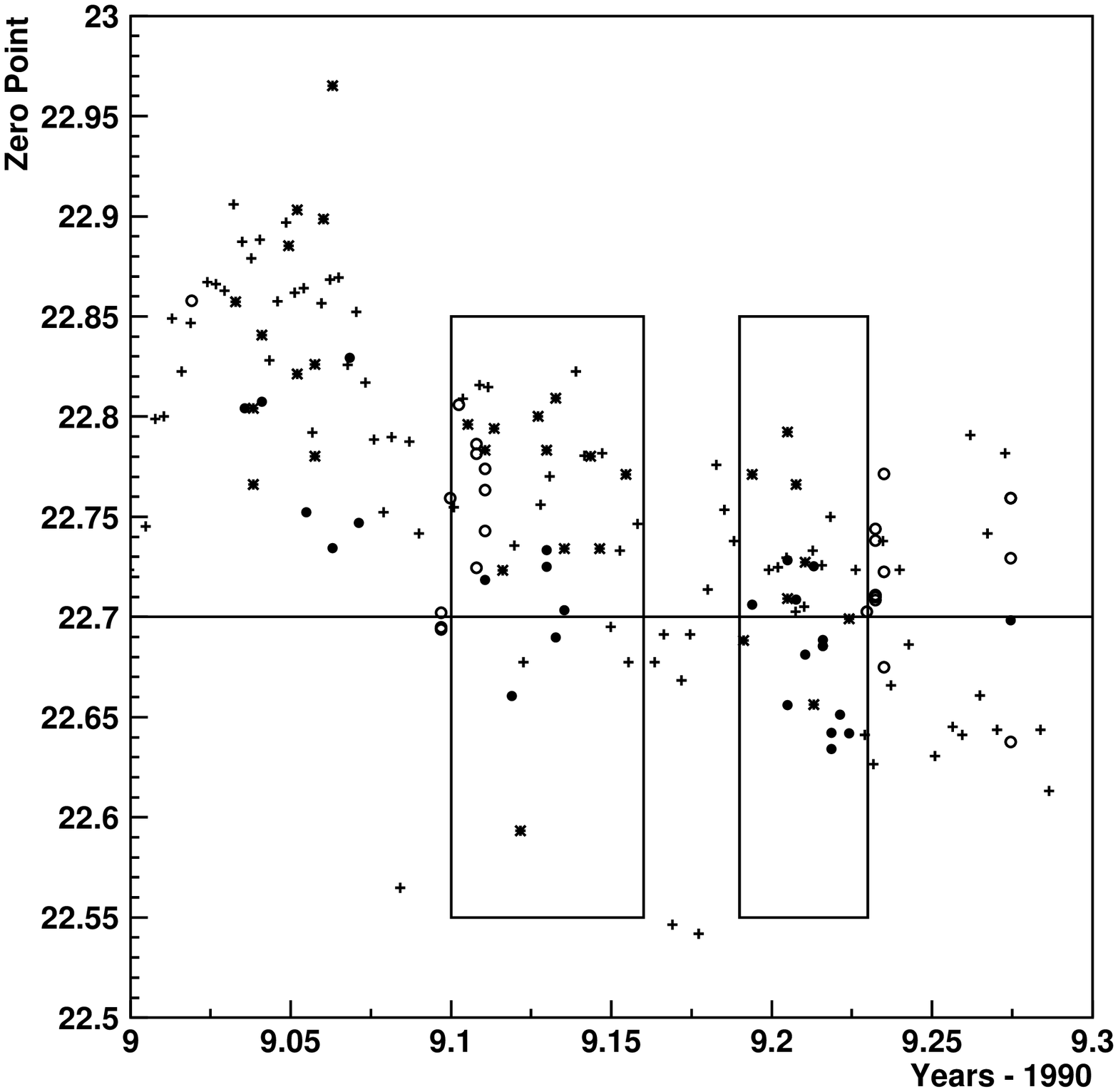}} 
\subfigure[Zoom on the 2000 supernova search: the mean zero point is
  5~\% lower than the one from 1999 searches (see box).]{
\label{fig:PZ00}
\includegraphics[width=8cm]{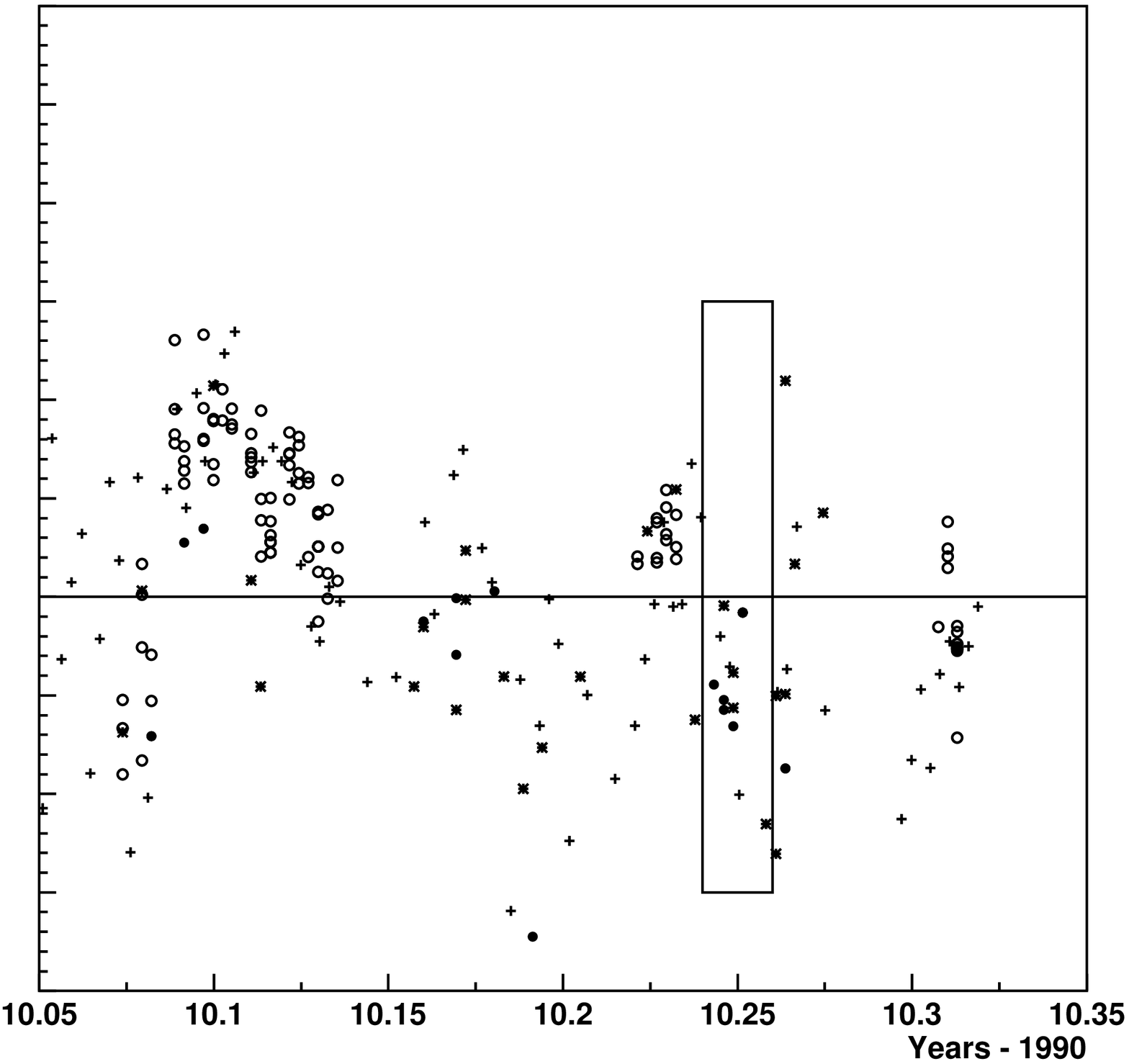}}
\caption{Evolution of the ``blue'' zero point as a function of
  time. Circles stand for the calibration using the Landolt catalog;
  black filled circles are for secondary calibrating stars from the
  supernova calibration of SCP 1999 set; crosses are absorption
  coefficients from EROS analysis of LMC and SMC stars, shifted to
  match the Landolt points.  The February 1999 search has been done
  between 02/04 and 02/27; the March 1999 search was performed between
  03/09 and 03/27; The April 2000 search took place between 03/27 and
  04/09. For these three campaigns the corresponding zero point is
  22.7 within a 5~\% range. The total uncertainty on this measurement
  is, from the scatter of the points, about $\pm$7~\%.}
\label{fig:PZ}
\end{center}
\end{figure*}
Zero points for the EROS observations are computed from standard stars
\citep{landolt1992} or secondary standard stars around some EROS
supernovae which were followed-up and analyzed by the SCP
\citep{regnault2001}. As the observations were always done close to
the meridian, the atmospheric absorption correction is included in the
zero point.  The evolution of the zero point as a function of time is
plotted in Fig.~\ref{fig:PZ}.  Within a 5~\% accuracy, it is the same
for the 1999 and the 2000 searches.  The color term is computed using
the SMC and LMC OGLE calibrated catalog \citep{udalski2000} in order
to have a larger lever arm: the Landolt catalog has the drawback to
have mainly red (87~\% have $B-V \gtrsim 0.3$) stars. This can be
overcome with the OGLE catalog of SMC and LMC stars, that EROS is
surveying too in its microlensing search program.  The result is
\begin{equation}
\widetilde{\mathcal{B}}_{\text{eros}} = V - 0.53 \cdot 
   (V-R) - 22.7.
\label{eqn:zpsne}
\end{equation}
The zero point above is obtained for one CCD of the EROS mosaic, the
seven other CCDs (that have different gains) are calibrated relative
to this one, using the sky background and various photometric
catalogs. The relative variation from CCD to CCD is of order of 10~\%.

\subsubsection{Zero point for the galaxies}

To calibrate the galaxy magnitudes we use the LCRS catalog which
provides a single isophotal $R_{\text{lcrs}}$ magnitude
\citep{lin1996}. We find:
\begin{equation}
\widetilde{\mathcal{B}}_{\text{eros}} = R_{\text{lcrs}} - 22.7.
\label{eqn:zpgal}
\end{equation}
\cite{lin1996} estimate that $R = R_{\text{lcrs}} - 0.25$.  The
distribution of $\widetilde{\mathcal{B}}_{\text{eros}} -
R_{\text{lcrs}}$ for the LCRS galaxies has a r.m.s. width less than
0.2 mag. This width is sufficiently narrow that it was not necessary
to refine eq. (\ref{eqn:zpgal}) with a color term.  The reason that
little precision is necessary is that the galaxy magnitudes are only
used, first, to derive the redshift probability distribution for each
galaxy and, second, to calculate the total galactic luminosity of the
sample to normalize the supernova rate.  In both cases, one
effectively averages over all galaxies in the sample so the relation
(\ref{eqn:zpgal}) is sufficiently accurate.

\subsubsection{Extinction}

Observations are always performed close to the meridian, so the
resulting atmospheric absorption can be considered constant and
included in the zero point.

The Galactic extinction is corrected field by field using the
\cite{schlegel1998} reddening map, assuming a standard Galactic
extinction law. The mean absorption is $A_V~\sim~0.12$ for the EROS
fields.

We do not correct for the supernova host galaxy extinction (or
intergalactic extinction) since the template light curves from
\cite{hamuy1996c} and \cite{riess1999} do not correct for it either.
We thus implicitly assume that the absorption in these two samples is
representative of supernovae in general.

\subsubsection{$K$-corrections}
\label{sec:kcorrection}

Since the supernovae are observed at different redshifts, their
spectra will move across the fixed passband used. In a given passband,
the sampled part of the spectrum thus depends on redshift and this
influences the measured flux.  This is taken into account by the
``$K$-correction''. The correction in the band $F$ is given by
\begin{equation}
K_F(t,z) = -2.5 \cdot \log \left[ 
\frac{\int_0^{+\infty}\frac{d\lambda_o}{1+z} F(\lambda_o) 
I\left(\frac{t}{1+z}, \frac{\lambda_o}{1+z}\right)}
{\int_0^{+\infty}d\lambda_o F(\lambda_o) I(t, \lambda_o)}\right],
\label{eqn:correctionK2}
\end{equation}
where $\lambda_o$ is the observed wavelength, $F$ is the
filter transmission function and $I$ is the normalized time-dependent
spectrum, which may depend on time $t$ -- like for supernovae. Note
that $K_F = 0$ when $F=1$, \textit{i.e.} when the passband covers the
whole spectrum.

For type~Ia supernovae they are calculated using a spectral
template\footnote{If we use the improved spectral template by
\cite{nobili2003} the resulting value of the rate is increased by
$\sim$ 1~\% which is negligible.}  \citep{nugent2002} and the shape of
the standard $V$ and $R$ filters (which are the bands used to
calibrate the EROS band, according to eq.~(\ref{eqn:zpsne})).  For
galaxies we use the approximation $K_{R}(z) = 2.5\cdot \log (1+z)$
which is good enough up to $z \sim 0.2$ \citep{poggianti1997}.

\section{Results}
\label{sec:result}

\begin{figure}[!htb]
\begin{center}
\includegraphics[width=8cm]{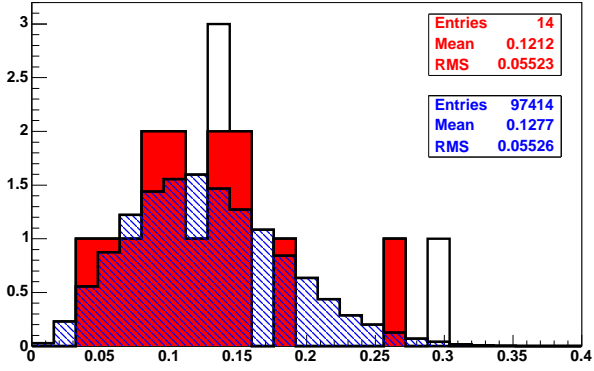}
\caption{Observed (red, filled -- upper box) and simulated (blue,
hatched -- lower box) redshift distribution. The two empty bins are
SNe for which the host galaxy magnitude is fainter than cuts
(\ref{eqn:cutgal99}) or (\ref{eqn:cutgal00}). A Kolmogorov-Smirnov
test gives a 85~\% probability of adequacy between the two
distributions.}
\label{fig:distzobs}
\end{center}
\end{figure}

\begin{figure}[!htb]
\begin{center}
\includegraphics[width=8cm]{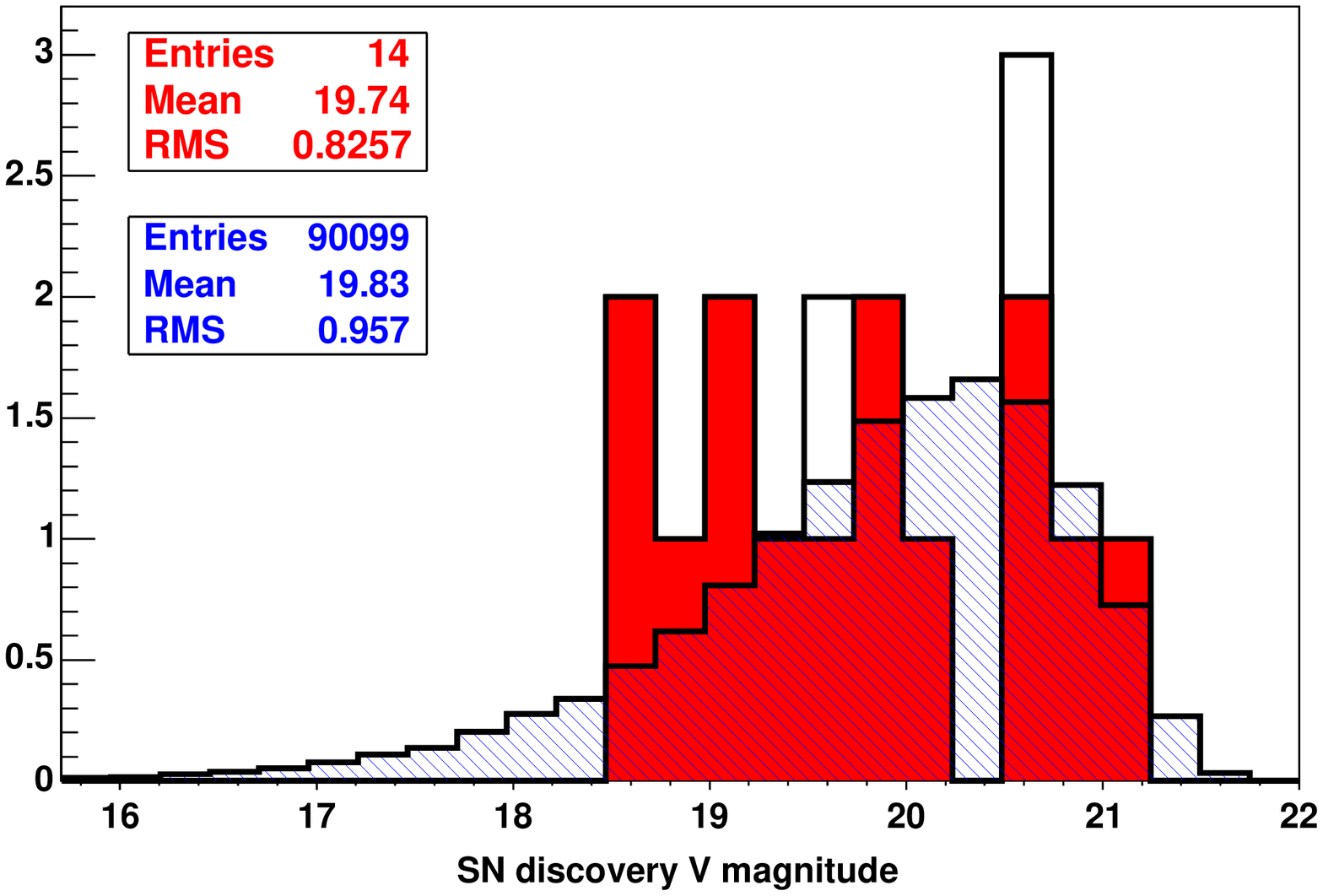}
\caption{Observed (red, filled -- upper box) and simulated (blue,
hatched -- lower box) discovery $V$ magnitude distribution. The two
empty bins are SNe for which the host galaxy magnitude is fainter than
cuts (\ref{eqn:cutgal99}) or (\ref{eqn:cutgal00}). A
Kolmogorov-Smirnov test gives a 62~\% probability of adequacy between
the two distributions.}
\label{fig:distmagobs}
\end{center}
\end{figure}

\begin{figure}[!htb]
\begin{center}
\includegraphics[width=8cm]{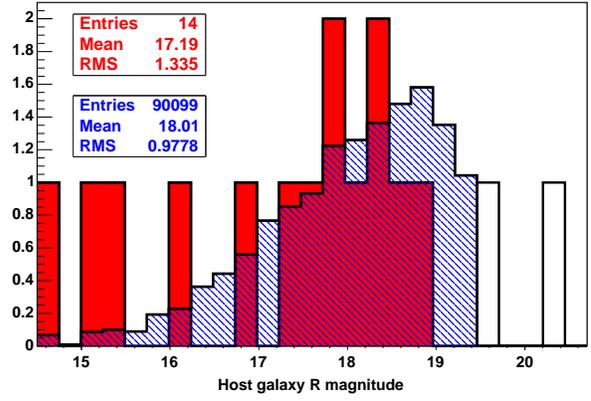}
\caption{Observed (red, filled -- upper box) and simulated (blue,
hatched -- lower box) host galaxy $R$ magnitude distribution. The two
empty bins are SNe for which the host galaxy magnitude is fainter than
cuts (\ref{eqn:cutgal99}) or (\ref{eqn:cutgal00}). A
Kolmogorov-Smirnov test gives a 23~\% probability of adequacy between
the two distributions.}
\label{fig:distgalmagobs}
\end{center}
\end{figure}

To check the validity of the Monte Carlo simulation we compare some
simulated distributions to the observed ones. Quantitative comparisons
are performed using a Kolmogorov-Smirnov (KS)
test. Figures~\ref{fig:distzobs} to \ref{fig:distgalmagobs} give
observed and simulated distributions for various parameters, namely
the redshift, the magnitude at detection and the host magnitude. The
similarity between the observed and the simulated distributions gives
confidence in the reliability of the integral computation. Considering
the limited statistics and the KS probabilities (85~\% for the
redshift -- Fig.~\ref{fig:distzobs}; 62~\% for the discovery magnitude
-- Fig.~\ref{fig:distmagobs}; 23~\% for the host galaxy magnitude --
Fig.~\ref{fig:distgalmagobs}) give no evidence for systematic
differences between theoretical and observed distributions.

\begin{figure}[!htb]
\begin{center}
\includegraphics[width=8cm]{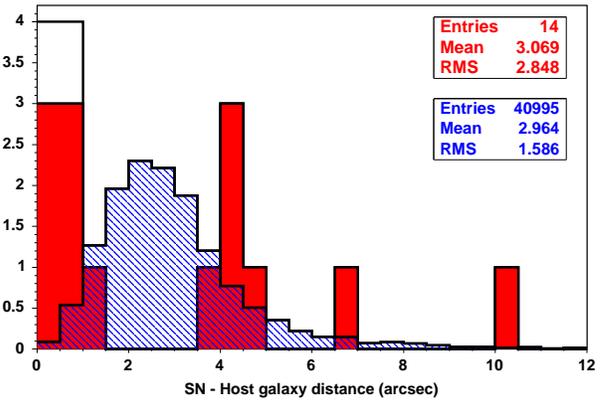}
\caption{Observed (red, filled -- upper box) and simulated (blue,
hatched -- lower box) distance to host distribution. The two empty
bins are SNe for which the host galaxy magnitude is fainter than cuts
(\ref{eqn:cutgal99}) or (\ref{eqn:cutgal00}). A Kolmogorov-Smirnov
test gives a 3~\% probability of adequacy between the two
distributions. The simulated distribution assumes that the type Ia
supernova rate is proportional to the galaxy surface brightness.}
\label{fig:distdistobs}
\end{center}
\end{figure}

The distribution of the distance between the supernova and the host
center is shown in Fig.~\ref{fig:distdistobs}.  The Monte Carlo
distribution was generated assuming that the supernova rate is
proportional to the local surface brightness.  The observed
distribution is more concentrated toward the center than the surface
brightness with the KS test giving a 3~\% probability for
compatibility.  This suggests supernovae do not have the same
distribution as the light though the limited statistics prevent any
firm conclusion.  At any rate, except far from galactic centers, our
detection efficiency is nearly independent of the distance to the host
center so the precise distribution of the distance has little effect
on the global efficiency.  All type of host galaxies are included in
the simulated distribution. The observed distribution confirms that we
are able to detect supernovae in the center of their host galaxies.

These checks being done, we are able to compute the rate, as the ratio
of the number of detected supernovae to the integral value calculated
as above.  From the 1999 search we have 12 type~Ia supernovae. But cut
(\ref{eqn:cutgal99}) on the galaxy host magnitude discards two events
(SN~1999bl and SN~1999bn). In 2000, four type~Ia supernovae were
detected. Only 14 events remain after the cut.  We can rewrite
eq.~(\ref{eqn:tauxgene}) as
\begin{equation}
\mathcal{R}_{\text{SN}} =
\frac{\mathcal{N}_{\text{tot}}}{\mathcal{I}_{\text{tot}}},
\label{eqn:tauxgene2}
\end{equation}
where $\mathcal{N}_{\text{tot}}$ is the total number of events
surviving the cuts, and $\mathcal{I}_{\text{tot}}$ is the
corresponding value of the integral. For the 1999 search, the integral
is $\mathcal{I}^{99} = 0.89 \times 10^{14}\ h_{70}^{-2}\
\text{L}_{\odot}^R\ \text{yr}$, while it is $\mathcal{I}^{00} = 0.50
\times 10^{14}\ h_{70}^{-2}\ \text{L}_{\odot}^R\ \text{yr}$ for the
2000 search. The combined value of the integral for the two searches
is the sum $\mathcal{I}_{\text{tot}} = \mathcal{I}^{99} +
\mathcal{I}^{00} = 1.39\times 10^{14}\ h_{70}^{-2}\
\text{L}_{\odot}^R\ \text{yr}$.  Assuming the number of observed
supernovae follows a Poisson law, with a confidence level of 68.3~\%,
one has $\mathcal{N}_{\text{tot}} = 14^{+4.83}_{-3.70}$.  This gives
$\mathcal{R}_{Ia}^R = 0.100^{+0.035}_{-0.027}\ h_{70}^2\
\text{SNu}_R$, where 1~SNu$_R$ is 1~supernova/$10^{10}\
\text{L}_{\odot}^R$/century.  The errors are only statistical at this
point.  The mean redshift for this value is given by the mean of the
simulated redshift distribution (Fig.~\ref{fig:distzobs}), that is
$\langle z\rangle \ = 0.128$. The calculated r.m.s. width of this
redshift distribution is 0.056.

\subsection{Systematic uncertainties}

We have identified four possible sources of systematics: the
calibration for both supernovae and galaxies, the cosmological model
and the assumed distributions of supernova luminosities and
light-curve shapes. For each of these parameters, various simulations
have been performed to quantify the corresponding impact on the rate.

\subsubsection{Calibration}

We estimate the uncertainty on the zero point of eq.~(\ref{eqn:zpsne})
to be of 0.07 mag (see Fig.~\ref{fig:PZ}). An increase
(resp. decrease) in the zero point of 7~\% decreases (resp. increases)
the rate by 6~\%.

Considering the galaxy calibration, eq.~(\ref{eqn:zpgal}), we estimate
an uncertainty on the zero point of about 0.1 mag (as combination of
statistical error in the ensemble zeropoint error along with a
possible systematic calibration error). An increase (resp. decrease)
of the zero point by 10~\% increases the rate by 7~\% (resp. decreases
the rate by 6~\%).  The net effect is lower than the 10~\% due to
luminosity alone, because the calibration for the redshift
distribution acts in the opposite way\footnote{This can be understood
in the following way: if the zero point is increased, the
corresponding $R$ magnitude increases for a given EROS flux. Then the
redshift distribution is shifted toward higher redshifts. This change
increases the value of the integral and thus decreases the rate.}.

\subsubsection{Cosmology}

The calculations and simulations have been done with the
($\Omega_{M_{\circ}},\ \Omega_{\Lambda_{\circ}}$) = (0.3,~0.7)
cosmological model. We checked that using the ($\Omega_{M_{\circ}},\
\Omega_{\Lambda_{\circ}}$) = (1,~0) model lowers the rate by about
1~\%. Only the probability density $p(z|m_g)$, the galaxy luminosities
and the supernovae magnitude (see eq.~(\ref{eqn:m1m2})), through the
luminosity distance, depend on the cosmological model. But while the
luminosity distance increases from (1,~0) model to (0.3,~0.7) model,
on the contrary, the $p$ distribution is shifted toward the lower
redshifts for a given apparent magnitude. The two effects cancel in
this redshift range, hence the very small dependence of the
cosmological parameters on the rate.

\subsubsection{Supernova diversity}
\label{sec:errorsnlf}

To take into account the fact that the distribution of supernovae
luminosities and light-curve shapes is currently uncertain due to a
lack of statistics, we parametrized the rate with the two main
variables describing the type~Ia variety distribution, the mean
absolute magnitude at maximum $\langle M_{\text{V}_{\text{max}}} + 5
\log h_{70}\rangle$, and the light curve shape parameter $\langle
\Delta m_{15}^B \rangle$ \citep{hamuy1996c} for the 17 light curves we
have used to derive the rate (see Fig.~\ref{fig:snelf}
and~\ref{fig:snelfdm15}). Then we compute the following empirical
relation:
\begin{equation}
\begin{split}
&\left(\frac{\mathcal{R}_{\text{Ia}}}{0.100\ h_{70}^2\
    \text{SNu}_R}\right) =  
\left(\frac{\langle \Delta m_{15}^B\rangle}{1.138} \right)^{0.65}\\ &
\quad \times \left[1 + 0.78\cdot (\langle M_{\text{V}_{\text{max}}} +
5 \log h_{70}\rangle + 18.55) \right],
\label{eqn:rateparam}
\end{split}
\end{equation}
for deviations up to $\sim$ 20~\% for $\langle
M_{\text{V}_{\text{max}}} + 5 \log h_{70}\rangle$ and up to 10~\% for
$\langle \Delta m_{15}^B\rangle$. If the whole type~Ia variety
distribution is shifted, either in luminosity or in light curve shape
parameter, as statistics increase with future experiments, our rate
value will be able to be adjusted accordingly to
eq.~(\ref{eqn:rateparam}).

The uncertainty on the mean of the peak luminosity distribution
(Fig.~\ref{fig:snelf}) of $\sim 0.96/\sqrt{17} = 0.23$ mag translates
into an uncertainty on the rate of $0.78 \times 0.23 = 0.18$. The
uncertainty on the mean value of $\Delta m_{15}^B$
(Fig.~\ref{fig:snelfdm15}), of $0.257/\sqrt{17} = 0.062$, translates
into an uncertainty on the rate of 4~\%. Then, the main source of
uncertainty on the rate comes from the mean value of the peak
luminosity distribution of supernovae.

If we remove the lowest luminosity supernova (SN~1996ai -- see
Fig.~\ref{fig:snelf}) from our reference sample, the rate is decreased
by 16~\%.

\subsubsection{Supernovae in dim galaxies}

Our search strategy requires an association to a host galaxy. This
prevents detecting supernovae appearing in low brightness galaxies
whose magnitude is beyond the detection threshold. For example, with
such a cut, the type~Ia SN~1999bw supernova, the host of which is a
very low brightness dwarf galaxy \citep{strolger2002}, would not have
been detected by EROS (this supernova had a $B$ magnitude of 16.9 at
maximum light, while its host has a $B$ magnitude of
24.2). \cite{gal-yam2003} also found two hostless type~Ia supernovae
in galaxy clusters. They argue that these events can be due to a
putative intergalactic star population.

Since we assume that the number of supernovae is proportional to the
host galaxy luminosity, the derived rate is independent of the galaxy
luminosity cut (eq.~(\ref{eqn:cutgal99}) and~(\ref{eqn:cutgal00}))
(apart from statistical fluctuation). Then such an eventuality is
included in our rate derivation as long as our working hypothesis is
valid. Moreover we remark that these events are very rare, most
probably representing a small fraction of events in the field.

\begin{table}[!htb]
\centering
\begin{tabular}{l@{}l|c}
Calibration: & \ SNe & $\pm 6\ \%$ \\
             & \ galaxies & $_{-6\ \%}^{+7\ \%}$ \\
 &  & \\
SNe light curve distribution: & \ $\langle \Delta m_{15}^B\rangle$ &
             $\pm 4\ \%$ \\  
                                   & \ $\langle M_{V_{\text{max}}}\rangle$ &
 $\pm 18\ \%$ \\  
\hline
\hline
Total (quadratic sum) &  & $\pm 21\ \%$\\
\end{tabular}
\caption{Summary of identified systematic errors. The resulting total
  systematic uncertainty can be compared to the statistical
  uncertainty: $^{+39\ \%}_{-23\ \%}$.}
\label{tab:errorbudget}
\end{table}

\subsection{The rate}

Table~\ref{tab:errorbudget} summarizes the error budget as discussed
above. Taking it into account gives for the rate, the value:
\begin{equation}
\mathcal{R}_{\text{Ia}}^{R} =
0.100^{+0.035+0.021}_{-0.027-0.021}\ h_{70}^{2}\ 
\text{SNu}_R,
\label{eqn:tauxtotR}
\end{equation}
where the first quoted error is statistical and the second
systematic. This number comes naturally in the $R$ band. However, the
SN rate is traditionally expressed per luminosity unit in the $B$
band.  We can convert this measurement into the $R$ band as:
\begin{equation}
\frac{\mathcal{R}_{\text{Ia}}^{B}}{\mathcal{R}_{\text{Ia}}^{R}} =
\frac{L_{\text{gal}}^R}{L_{\odot}^R} \cdot
\frac{L_{\odot}^B}{L_{\text{gal}}^B} =
10^{-0.4\cdot[(B-R)_{\odot} - (B-R)_{\text{gal}}]}.
\end{equation}
Using $(B-R)_{\odot} = 1.06$ and $(B-R)_{\text{gal}} = 1.3$ as
the mean of the whole galaxy sample of \cite{grogin1999}; we take
$\pm 0.1$ mag as a possible systematic between our galaxy sample and
theirs. Then, we have
\begin{equation}
\frac{L_{\text{gal}}^R}{L_{\odot}^R} \cdot
\frac{L_{\odot}^B}{L_{\text{gal}}^B} = 1.25 \pm 0.11,
\label{eqn:r2b}
\end{equation}
which gives: 
\begin{equation}
\mathcal{R}_{\text{Ia}}^{B} = 0.125^{+0.044+0.028}_{-0.034-0.028} \
h_{70}^{2}\ \text{SNu},
\label{eqn:tauxtotB}
\end{equation}
where the uncertainty which appears in the right hand side of
eq.~(\ref{eqn:r2b}) has been added quadratically to the systematic
error.
 
To obtain the rate per comoving volume unit, we multiply the value in
eq.~(\ref{eqn:tauxtotB}) by the mean galaxy luminosity density. The
2dF Galaxy Redshift Survey \citep{cross2001} provides a value at a
mean redshift of 0.1, in the $B$ band and for a (1,~0) cosmological
model, in agreement with the one obtained by the ESO Slice Project
\citep{zucca1997}. We translate the 2dF value to a (0.3,~0.7)
model\footnote{Note that $j \propto \mathcal{D}^2/\mathcal{V}$
where $\mathcal{D}$ is the luminosity distance and $\mathcal{V} =
dV/(dz d\Omega)$ is the comoving volume element (see \textit{e.g.}
\cite{carroll1992}, eq.~(26)).}  ($\left. j^{(0.3,~0.7)} /
j^{(1,~0)}\right|_{z=0.1} = 0.91$): $j_B~=~(1.59 \pm 0.13)
\times 10^8\ h_{70}\ \text{L}_{\odot}\ \text{Mpc}^{-3}$. The supernova
rate (\ref{eqn:tauxtotB}) can then be expressed as:
\begin{equation}
\mathcal{R}_{\text{Ia}} = (1.99^{+0.70+0.47}_{-0.54-0.47})\times
10^{-5}\ h_{70}^3\  \text{yr}^{-1} \text{Mpc}^{-3}.
\label{eqn:tauxtotBvol}
\end{equation}
Here, the uncertainty on the luminosity density function has been added
quadratically to the systematic error bar. 

\section{Discussion}
\label{sec:discussion}

\subsection{Comparison with other measurements}

Table~\ref{tab:otheresults} shows other published results for the
type~Ia supernova rate at various redshifts. Our measurement compares
well with other measurements in the nearby universe
\citep{cappellaro1999,hardin2000,madgwick2003}. It is slightly lower
but compatible within error bars. We note that our SN search strategy
and our supernova sample are very different from those of
\cite{cappellaro1999} and \cite{madgwick2003}. The former used a
sample of nearby supernovae discovered either by eye or on
photographic plates which introduced large systematic errors in the
rate. The latter used a new search technique based on galaxy spectra:
supernovae are discovered in subtracting 116\,000 galaxy spectra from
the Sloan Digital Sky Survey by a eigenbasis of 20 ``unpolluted''
galaxy spectra. The method is very promising, especially in deriving
the rate as a function of host galaxy properties.

On the other hand our rate is lower than the distant \cite{pain2002}
value, by over one standard deviation, either for the rate as a
function of galaxy luminosity or as a function of comoving volume.
{\setlength{\extrarowheight}{4pt}
\begin{table*}[!ht]
\centering
\begin{tabular}{lllccl}
\hline
\hline
      & \multicolumn{2}{c}{$R_{\textrm{SN Ia}}$} 
&  &  &  \\
\cline{2-3}
\raisebox{2.5ex}[0cm][0cm]{$\langle z\rangle$}  & ($h_{70}^2$ SNu) & 
($10^{-5}\ h_{70}^3\
\text{Mpc}^{-3}\ \text{yr}^{-1}$) & 
\raisebox{2.5ex}[0cm][0cm]{($\Omega_{M_{\circ}}, 
\Omega_{\Lambda_{\circ}}$)} & \raisebox{2.5ex}[0cm][0cm]{SNe nb} &
\raisebox{2.5ex}[0cm][0cm]{author} \\ 
%\hline
\hline
$\sim$ 0    & $0.18\pm0.05$ & 2.8$\pm$0.9 & & 70 &
\cite{cappellaro1999}$^a$ \\
0.098 & 0.196 $\pm$ 0.098 & $3.12 \pm 1.58$ &  & 19 &
\cite{madgwick2003}$^a$ \\  
\mathversion{bold}
\textbf{0.13} & $\mathbf{0.125^{+0.044+0.028}_{-0.034-0.028}}$  &
$\mathbf{1.99^{+0.70+0.47}_{-0.54-0.47}}$  & \textbf{(0.3, 0.7)} &
  \textbf{14} &  
\textbf{this work} \\ 
\mathversion{normal}
0.14 & $0.22^{+0.17+0.06}_{-0.10-0.03}$ &
$3.43^{+2.7+1.1}_{-1.6-0.6} $& (0.3, 0.7) & 4 &
\cite{hardin2000}$^{a, b}$ \\
$0.25^{+0.12}_{-0.07}$ & $0.20^{+0.82}_{-0.19}$ & & (0.3, 0.7) & 1 &
\cite{gal-yam2002}$^c$\\  
0.38  & $0.40^{+0.26+0.18}_{-0.18-0.12}$ & & (1.0, 0.0) & 3 &
\cite{pain1996} \\ 
0.46  & & $4.8\pm 1.7$ & (0.3, 0.7) & 8 &
\cite{tonry2003} \\ 
0.55 & $0.28^{+0.05+0.05}_{-0.04-0.04}$ &
  $5.25^{+0.96+1.10}_{-0.86-1.06}$ & (0.3, 0.7) & 38 & 
\cite{pain2002} \\ 
0.55 & $0.46^{+0.08+0.07}_{-0.07-0.0.7}$ &
$11.1^{+2.0+1.1}_{-1.8-1.1}$ & (1.0, 0.0) & 38 & 
\cite{pain2002} \\ 
$0.90^{+0.37}_{-0.07}$ & $0.40^{+1.21}_{-0.38}$ & & (0.3, 0.7) & 5 &
\cite{gal-yam2002}$^c$\\  
%\hline
\end{tabular}
\caption{Comparison with other published restframe type~Ia supernova
explosion rate measurements. These rates are given together with the
mean redshift of the observed SNe and the cosmological model assumed
(especially for distant values). The fifth column gives the number of
supernovae from which the rate is computed. \textsc{Notes}: $a$)~the
value per comoving volume unit is derived from the SNu value using the
2dF luminosity density (the error on the luminosity density is added
quadratically to the systematic uncertainty if distinct);
$b$)~\protect\cite{hardin2000} computed the rate using a
($\Omega_{M_{\circ}}, \Omega_{\Lambda_{\circ}}$)~=~(0.3,~0.0)
cosmological model. Nevertheless no important difference from the
(0.3,~0.7) model is expected (less than 10\%);
$c$)~\protect\cite{gal-yam2002} have measured the rate in galaxy
clusters, with could be different from the rate in the field; they do
not give a value per unit of comoving volume.}
\label{tab:otheresults}
\end{table*}}

\subsection{Evolution?}

Our result combined with the SCP measurement at higher redshift
\citep{pain2002} suggests models with an evolution of the rate, either
in SNu or per unit of comoving volume, between $z = 0.13$ and $z =
0.55$. We can parametrize such evolution as a power-law:
$\mathcal{R}_{\text{Ia}}(z) = \mathcal{R}_{\text{Ia}}(0)\cdot
(1+z)^{\alpha}$ where $\alpha$ is a restframe evolution index. Using
the present result and the \cite{pain2002} value per unit of comoving
volume, we obtain $\alpha_v = 3.1^{+1.6}_{-1.4}$. The same parameter
computed for the rate expressed in SNu is $\alpha_l =
2.6^{+1.5}_{-1.3}$. The difference, $\alpha_v - \alpha_l = 0.5\pm
0.1$, simply reflects the different luminosity densities adopted by us
and \cite{pain2002}, equivalent to an evolving luminosity density with
$\alpha_{j_B}^{\text{SNe}} = 0.5\pm 0.1$. In fact, the luminosity
density most likely evolves a bit faster. Combining the data of
\cite{lilly1996} with that of \cite{cross2001} and imposing a
(0.3,~0.7) cosmological model, the evolution of the luminosity density
corresponds to $\alpha_{j_B}^{\text{Lilly}} = 0.73\pm0.51$. If
\cite{pain2002} had adopted a luminosity density more in line with
this value of $\alpha_{j_B}$ then we would have deduced $\alpha_l =
2.4^{+1.7}_{-1.5}$ by combining our data with that of \cite{pain2002}.
This number is still significantly greater than zero.

The value of $\alpha_v = 3.1^{+1.6}_{-1.4}$ derived by combining our
data with that of \cite{pain2002} is consistent with the value
$\alpha_v = 0.8\pm 1.6$ derived using only the \cite{pain2002} data.
\begin{figure}[!htb]
\begin{center}
\includegraphics[width=8.5cm]{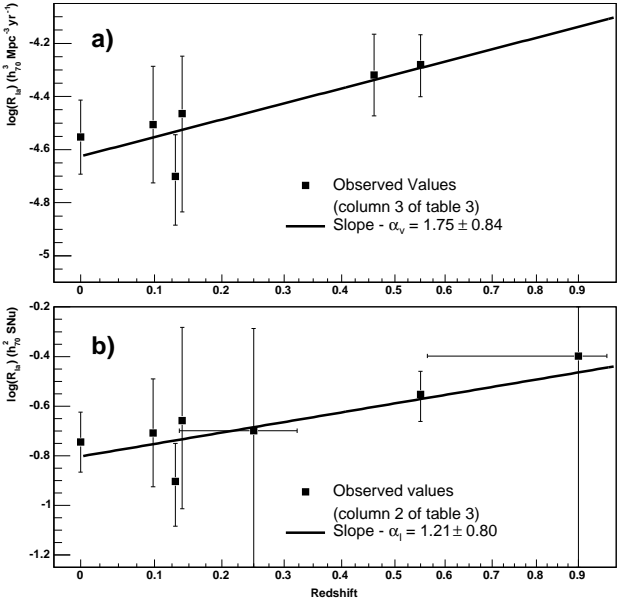}
\end{center}
\caption{Derivation of the evolution index $\alpha$ using all
available SNIa rate measurements (see Table~\ref{tab:otheresults}),
within the (0.3,~0.7) cosmological model.}
\label{fig:fitrates}
\end{figure}
Nevertheless, fitting all available rate measurements either in SNu or
per comoving volume unit, provides a lower evolution index (see
Fig.~\ref{fig:fitrates}): $\alpha_v = 1.75 \pm 0.84$
and $\alpha_l = 1.21\pm 0.80$.

\begin{figure*}[!htb]
\begin{center}
\includegraphics[width=17cm]{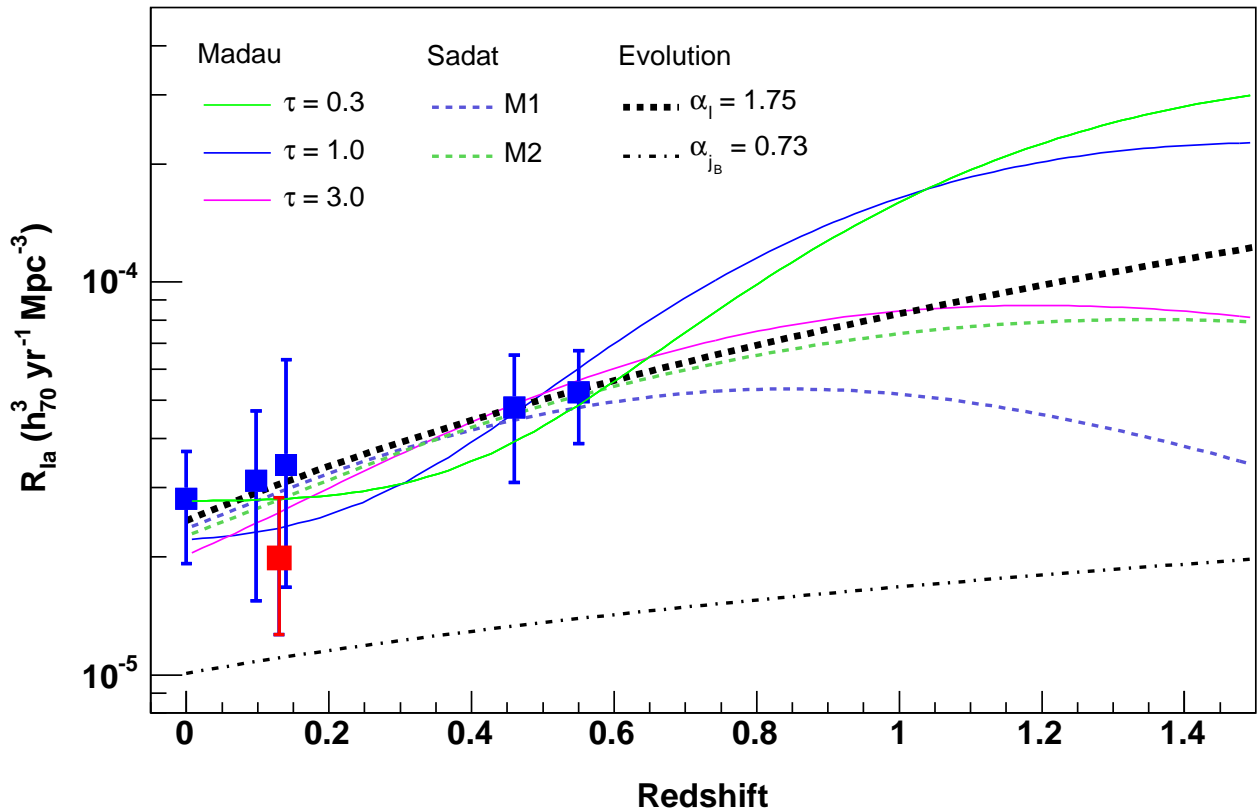}
\end{center}
\caption{Explosion rate evolution per unit of time and comoving
volume. The set of curves labeled ``Madau'' stands for the models by
\protect\cite{madau1998}, computed within the hierarchical clustering
galaxy formation scenario; $\tau$ is the time interval between the
white dwarf formation and the supernova explosion (in Gyr). The curves
labeled ``Sadat'' stand for the \protect\cite{sadat1998} models, using
two different star formation rate evolution models (M1 and M2, the
latter accounts for a possible dust extinction correction).  From left
to right are the measured values by \protect\cite{cappellaro1999},
\protect\cite{madgwick2003}, this work, \protect\cite{hardin2000},
\protect\cite{tonry2003} and \protect\cite{pain2002}. We plot also the
redshift evolution, $(1+z)^{1.75}$, corresponding to the best fit
using the six measured values.  The effect of the evolution of the
galaxy luminosity density ($\alpha_{j_B}$) is shown as well since it
is included in the rate per comoving volume unit evolution.  All data
points and model curves have been computed with the (0.3,~0.7)
cosmological model, and $h_{70} = 1$. The normalisation of models is
set by adjusting them to the data.}
\label{fig:snratemad}
\end{figure*}

Models such as those by \cite{madau1998} or by \cite{sadat1998} can
fit the observations (Fig.~\ref{fig:snratemad}). In the
\cite{madau1998} models, the evolution depends not only on the time
delay between the white dwarf formation and the supernova
explosion\footnote{According to \cite{dahlen1999}, the $\tau = 0.3$
Gyr model corresponds to the double degenerate progenitor scenario,
while the $\tau = 1$ Gyr model simulates the single degenerate -- or
cataclysmic -- progenitor scenario; the $\tau = 3$ Gyr expands the
range to cover all likely models.}, but also on the galaxy formation
scenario, especially at high redshift. In these models the supernova
rate follows roughly a $(1+z)^{\alpha_v}$ evolution with $\alpha_v
\gtrsim 1.8$ ($\tau = 0.3$), $\alpha_v \sim 2.4$ ($\tau = 3$) up to
$\alpha_v \sim 2.8$ ($\tau = 1$). The \cite{sadat1998} models depend
on the choice of star formation rate evolution model and on dust
extinction. They predict $\alpha_v \sim 1.4$ for ``M1'' model and
$\alpha_v \sim 1.8$ for the ``M2'' model.

Too few data points exist up to now in order to constrain accurately
many parameters from the models, such as the galaxy formation scenario
\citep{madau1998} (only good quality data at redshift $\gtrsim 1.5$
could eventually distinguish between the two favored galaxy formation
scenarios), the progenitor model \citep{ruiz-lapuente1998},
environmental effects like the metallicity \citep{kobayashi2000} or
the cosmology \citep{dahlen1999}. 

\subsection{Conclusion}

We have measured the type~Ia supernova explosion rate at a redshift of
0.13. Combined with other measurements at different redshifts, it
suggests a slight evolution of the rate with the
redshift. Nevertheless, the scatter and the still large error bars of
the current measurements do not allow us to make definite conclusions
about it.

Future supernova searches will shed more light on the rate
evolution. Among them, two dedicated supernova searches, are well
suited to improve and measure the evolution of the explosion rate. The
\textit{Nearby Supernova Factory} \citep{aldering2002} is a nearby ($z
\lesssim 0.1$) supernova search using two dedicated telescopes, one
for the search and one for the follow-up of discovered events using an
integral field spectrometer. It will be very helpful to measure the
local supernova rate using a very homogeneous set of a few hundred
events, and it will also help reducing the uncertainties on the
supernova light curve distribution.  Another supernova search in
progress uses the wide field \textsc{Megacam} camera on the
CFHT\footnote{http://snls.in2p3.fr}. Supernovae are discovered in a
rolling search mode up to $z \lesssim 0.9$; it will be possible to use
them for measuring the evolution of the explosion rate for type~Ia
supernovae at a 15~\% level of statistical uncertainty (per bin of 0.1
in redshift). Concerning other supernova searches well suited to
compute the explosion rate, we can quote the Sloan Digital Sky Survey
performing a spectroscopic detection of nearby supernovae, as already
mentioned \citep{madgwick2003}; the Lick
Observatory and Tenagra Observatory Supernova
Searches\footnote{http://astron.berkeley.edu/$\sim$bait/lotoss.html}
detecting nearby event using automatic telescopes; at intermediate
redshifts ($z \sim 0.4$), the ESO Supernova
Search\footnote{http://web.pd.astro.it/supern/esosearch/} is designed
to search for supernovae in order to derive the rate; in the distant
Universe, the Great Observatories Origins Deep Survey Treasury
program\footnote{http://www.stsci.edu/science/goods/} is searching for
very distant type Ia supernovae using the Hubble Space Telescope.

\begin{acknowledgements}
We thank R. Pain, E. Cappellaro and G. Altavilla for helpful
discussions. We also thank R. Sadat for providing ascii files of her
models, P. Madau for quick replies to questions about his models and
S. Lilly for discussions about the evolution of galaxy luminosity
density. We are grateful to the referee for perceptive
questions and comments that improved the manuscript.

This work was supported in part by the Director, Office of Science,
Office of High Energy and Nuclear Physics, of the U.S. Department of
Energy under Contract No.  DE-AC03-76SF000098.

The observations described in this paper were based in part on
observations using the CNRS/INSU Marly telescope at the European
Southern Observatory, La Silla, Chile; in part using the Nordic
Optical Telescope, operated on the island of La Palma jointly by
Denmark, Finland, Iceland, Norway, and Sweden, in the Spanish
Observatorio del Roque de los Muchachos of the Instituto de
Astrofisica de Canarias; in part using the Apache Point Observatory
3.5-meter telescope, which is owned and operated by the Astrophysical
Research Consortium; in part using telescopes at Lick Observatory,
which is owned and operated by the University of California; and in
part using telescopes at the National Optical Astronomy Observatories,
which are operated by the Association of Universities for Research in
Astronomy, Inc., under a cooperative agreement with the United States
National Science Foundation.
\end{acknowledgements}

% for the bibliography at the end
\bibliographystyle{aa} % style aa.bst
\bibliography{biblio}

\begin{thebibliography}{46}
\expandafter\ifx\csname natexlab\endcsname\relax\def\natexlab#1{#1}\fi

\bibitem[{{Afonso} {et~al.}(2003{\natexlab{a}}){Afonso}, {Albert}, {Alard},
  {Andersen}, {Ansari}, {Aubourg}, {Bareyre}, {Bauer}, {Beaulieu}, {Blanc},
  {Bouquet}, {Char}, {Charlot}, {Couchot}, {Coutures}, {Derue}, {Ferlet},
  {Fouqu{\' e}}, {Glicenstein}, {Goldman}, {Gould}, {Graff}, {Gros},
  {Haissinski}, {Hamadache}, {Hamilton}, {Hardin}, {de Kat}, {Kim}, {Lasserre},
  {LeGuillou}, {Lesquoy}, {Loup}, {Magneville}, {Mansoux}, {Marquette},
  {Maurice}, {Maury}, {Milsztajn}, {Moniez}, {Palanque-Delabrouille},
  {Perdereau}, {Pr{\' e}vot}, {Regnault}, {Rich}, {Spiro}, {Tisserand},
  {Vidal-Madjar}, {Vigroux}, \& {Zylberajch}}]{afonso2003a}
{Afonso}, C., {Albert}, J.~N., {Alard}, C., {et~al.} 2003{\natexlab{a}}, \aap,
  404, 145, (The EROS collaboration)

\bibitem[{{Afonso} {et~al.}(2003{\natexlab{b}}){Afonso}, {Albert}, {Andersen},
  {Ansari}, {Aubourg}, {Bareyre}, {Beaulieu}, {Blanc}, {Charlot}, {Couchot},
  {Coutures}, {Ferlet}, {Fouqu{\' e}}, {Glicenstein}, {Goldman}, {Gould},
  {Graff}, {Gros}, {Haissinski}, {Hamadache}, {de Kat}, {Lasserre}, {Le
  Guillou}, {Lesquoy}, {Loup}, {Magneville}, {Marquette}, {Maurice}, {Maury},
  {Milsztajn}, {Moniez}, {Palanque-Delabrouille}, {Perdereau}, {Pr{\' e}vot},
  {Rahal}, {Rich}, {Spiro}, {Tisserand}, {Vidal-Madjar}, {Vigroux}, \&
  {Zylberajch}}]{afonso2003b}
{Afonso}, C., {Albert}, J.~N., {Andersen}, J., {et~al.} 2003{\natexlab{b}},
  \aap, 400, 951, (The EROS collaboration)

\bibitem[{{Aldering} {et~al.}(2002){Aldering}, {Adam}, {Antilogus}, {Astier},
  {Bacon}, {Bongard}, {Bonnaud}, {Copin}, {Hardin}, {Henault}, {Howell},
  {Lemonnier}, {Levy}, {Loken}, {Nugent}, {Pain}, {Pecontal}, {Pecontal},
  {Perlmutter}, {Quimby}, {Schahmaneche}, {Smadja}, \&
  {Wood-Vasey}}]{aldering2002}
{Aldering}, G., {Adam}, G., {Antilogus}, P., {et~al.} 2002, in Survey and Other
  Telescope Technologies and Discoveries. Edited by Tyson, J. Anthony; Wolff,
  Sidney. Proceedings of the SPIE, Volume 4836, pp. 61-72 (2002)., 61--72

\bibitem[{Bauer \& de~Kat(1997)}]{bauer1997}
Bauer, F. \& de~Kat, J. 1997, in Optical Detectors for Astronomy, held at ESO,
  Garching, october 8-10, 1996, ed. J.~W. Beletic \& P.~Amico, (The EROS
  collaboration)

\bibitem[{Bertin \& Arnouts(1996)}]{bertin1996}
Bertin, E. \& Arnouts, S. 1996, \aaps, 117, 393

\bibitem[{Bertin \& Dennefeld(1997)}]{bertin1997}
Bertin, E. \& Dennefeld, M. 1997, \aap, 317, 43

\bibitem[{Cappellaro {et~al.}(1999)Cappellaro, Evans, \&
  Turatto}]{cappellaro1999}
Cappellaro, E., Evans, R., \& Turatto, M. 1999, \aap, 351, 459

\bibitem[{{Cappellaro} {et~al.}(1993){Cappellaro}, {Turatto}, {Benetti},
  {Tsvetkov}, {Bartunov}, \& {Makarova}}]{cappellaro1993}
{Cappellaro}, E., {Turatto}, M., {Benetti}, S., {et~al.} 1993, \aap, 273, 383

\bibitem[{Carroll {et~al.}(1992)Carroll, Press, \& Turner}]{carroll1992}
Carroll, S.~M., Press, W.~H., \& Turner, E.~L. 1992, \araa, 30, 499

\bibitem[{Contardo {et~al.}(2000)Contardo, Leibundgut, \& Vacca}]{contardo2000}
Contardo, G., Leibundgut, B., \& Vacca, W.~D. 2000, \aap, 359, 876

\bibitem[{{Cousins}(1976)}]{cousins1976}
{Cousins}, A.~W.~J. 1976, \memras, 81, 25

\bibitem[{{Cross} {et~al.}(2001){Cross}, {Driver}, {Couch}, {Baugh},
  {Bland-Hawthorn}, {Bridges}, {Cannon}, {Cole}, {Colless}, {Collins},
  {Dalton}, {Deeley}, {De Propris}, {Efstathiou}, {Ellis}, {Frenk},
  {Glazebrook}, {Jackson}, {Lahav}, {Lewis}, {Lumsden}, {Maddox}, {Madgwick},
  {Moody}, {Norberg}, {Peacock}, {Peterson}, {Price}, {Seaborne}, {Sutherland},
  {Tadros}, \& {Taylor}}]{cross2001}
{Cross}, N., {Driver}, S.~P., {Couch}, W., {et~al.} 2001, \mnras, 324, 825

\bibitem[{{Dahl{\' e}n} \& {Fransson}(1999)}]{dahlen1999}
{Dahl{\' e}n}, T. \& {Fransson}, C. 1999, \aap, 350, 349

\bibitem[{{Gal-Yam} {et~al.}(2003){Gal-Yam}, {Maoz}, {Guhathakurta}, \&
  {Filippenko}}]{gal-yam2003}
{Gal-Yam}, A., {Maoz}, D., {Guhathakurta}, P., \& {Filippenko}, A.~V. 2003,
  \aj, 125, 1087

\bibitem[{{Gal-Yam} {et~al.}(2002){Gal-Yam}, {Maoz}, \& {Sharon}}]{gal-yam2002}
{Gal-Yam}, A., {Maoz}, D., \& {Sharon}, K. 2002, \mnras, 332, 37

\bibitem[{{Grogin} \& {Geller}(1999)}]{grogin1999}
{Grogin}, N.~A. \& {Geller}, M.~J. 1999, \aj, 118, 2561

\bibitem[{Hamuy {et~al.}(1996)Hamuy, Phillips, Suntzeff, Schommer, Maza,
  Antezan, Wischnjewsky, Valladares, Muena, Gonzales, Aviles, Wells, Smith,
  Navarrete, Covarrubias, Williger, Walker, Layden, Elias, Baldwin, Hernandez,
  Tirado, Ugarte, Elston, Saavedra, Barrientos, Costa, Lira, Ruiz, Anguita,
  Gomez, Ortiz, della Valle, Danziger, Storm, Kim, Bailyn, Rubenstein, Tucker,
  Cersosimo, Mendez, Siciliano, Sherry, Chaboyer, Koopmann, Geisler,
  Sarajedini, Dey, Tyson, Rich, Gal, Lamontagne, Caldwell, Guhathakurta,
  Phillips, Szkody, Prosser, Ho, McMahan, Baggley, Cheng, Havlen, Wakamatsu,
  Janes, Malkan, Baganoff, Seitzer, Shara, Sturch, Hesser, Hartig, Hughes,
  Welch, Williams, Ferguson, Francis, French, Bolte, Roth, Odewahn, Howell, \&
  Krzeminski}]{hamuy1996c}
Hamuy, M., Phillips, M.~M., Suntzeff, N.~B., {et~al.} 1996, \aj, 112, 2408

\bibitem[{Hardin {et~al.}(2000)Hardin, Afonso, Alard, Albert, Amadon, Andersen,
  Ansari, Aubourg, Bareyre, Bauer, Beaulieu, Blanc, Bouquet, Char, Charlot,
  Couchot, Coutures, Derue, Ferlet, Glicenstein, Goldman, Gould, Graff, Gros,
  Haissinski, Hamilton, de~Kat, Kim, Lasserre, Lesquoy, Loup, Magneville,
  Mansoux, Marquette, Maurice, Milsztajn, Moniez, Palanque-Delabrouille,
  Perdereau, Pr\'evot, Regnault, Rich, Spiro, Vidal-Madjar, Vigroux, \&
  Zylberajch}]{hardin2000}
Hardin, D., Afonso, C., Alard, C., {et~al.} 2000, \aap, 362, 419, (The EROS
  Collaboration)

\bibitem[{{Howell} {et~al.}(2000){Howell}, {Wang}, \& {Wheeler}}]{howell2000}
{Howell}, D.~A., {Wang}, L., \& {Wheeler}, J.~C. 2000, \apj, 530, 166

\bibitem[{{Johnson}(1965)}]{johnson1965}
{Johnson}, H.~L. 1965, Communications of the Lunar and Planetary Laboratory, 3,
  73

\bibitem[{{Kobayashi} {et~al.}(2000){Kobayashi}, {Tsujimoto}, \&
  {Nomoto}}]{kobayashi2000}
{Kobayashi}, C., {Tsujimoto}, T., \& {Nomoto}, K. 2000, \apj, 539, 26

\bibitem[{Landolt(1992)}]{landolt1992}
Landolt, A.~U. 1992, \aj, 104, 340

\bibitem[{{Lasserre} {et~al.}(2000){Lasserre}, {Afonso}, {Albert}, {Andersen},
  {Ansari}, {Aubourg}, {Bareyre}, {Bauer}, {Beaulieu}, {Blanc}, {Bouquet},
  {Char}, {Charlot}, {Couchot}, {Coutures}, {Derue}, {Ferlet}, {Glicenstein},
  {Goldman}, {Gould}, {Graff}, {Gros}, {Ha{\i}ssinski}, {Hamilton}, {Hardin},
  {de Kat}, {Kim}, {Lesquoy}, {Loup}, {Magneville}, {Mansoux}, {Marquette},
  {Maurice}, {Milsztajn}, {Moniez}, {Palanque-Delabrouille}, {Perdereau},
  {Pr{\' e}vot}, {Regnault}, {Rich}, {Spiro}, {Vidal-Madjar}, {Vigroux},
  {Zylberajch}, \& {The EROS collaboration}}]{lasserre2000}
{Lasserre}, T., {Afonso}, C., {Albert}, J.~N., {et~al.} 2000, \aap, 355, L39,
  (The EROS collaboration)

\bibitem[{{Lilly} {et~al.}(1996){Lilly}, {Le Fevre}, {Hammer}, \&
  {Crampton}}]{lilly1996}
{Lilly}, S.~J., {Le Fevre}, O., {Hammer}, F., \& {Crampton}, D. 1996, \apjl,
  460, 1

\bibitem[{Lin {et~al.}(1996)Lin, Kirshner, Shectman, Landy, Oemler, \&
  Tucker}]{lin1996}
Lin, H., Kirshner, R.~P., Shectman, S.~A., {et~al.} 1996, \apj, 464, 60

\bibitem[{Livio(2001)}]{livio2001}
Livio, M. 2001, in The greatest Explosions since the Big Bang: Supernovae and
  Gamma-Ray Bursts, held 3-6 May, 1999 at Space Telescope Science Institute,
  Baltimore, MD, ed. M.~Livio, N.~Panagia, \& K.~Sahu (Cambridge University
  Press), 334, astro-ph/0005344

\bibitem[{{Madau} {et~al.}(1998){Madau}, {della Valle}, \&
  {Panagia}}]{madau1998}
{Madau}, P., {della Valle}, M., \& {Panagia}, N. 1998, MNRAS, 297, 17

\bibitem[{{Madgwick} {et~al.}(2003){Madgwick}, {Hewett}, {Mortlock}, \&
  {Wang}}]{madgwick2003}
{Madgwick}, D.~S., {Hewett}, P.~C., {Mortlock}, D.~J., \& {Wang}, L. 2003,
  \apjl, 599, L33

\bibitem[{{Nobili} {et~al.}(2003){Nobili}, {Goobar}, {Knop}, \&
  {Nugent}}]{nobili2003}
{Nobili}, S., {Goobar}, A., {Knop}, R., \& {Nugent}, P. 2003, \aap, 404, 901

\bibitem[{Nugent {et~al.}(2002)Nugent, Kim, \& Perlmutter}]{nugent2002}
Nugent, P., Kim, A., \& Perlmutter, S. 2002, PASP, 114, 803

\bibitem[{{Pain} {et~al.}(2002){Pain}, {Fabbro}, {Sullivan}, {Ellis},
  {Aldering}, {Astier}, {Deustua}, {Fruchter}, {Goldhaber}, {Goobar}, {Groom},
  {Hardin}, {Hook}, {Howell}, {Irwin}, {Kim}, {Kim}, {Knop}, {Lee}, {Lidman},
  {McMahon}, {Nugent}, {Panagia}, {Pennypacker}, {Perlmutter}, {Ruiz-Lapuente},
  {Schahmaneche}, {Schaefer}, \& {Walton}}]{pain2002}
{Pain}, R., {Fabbro}, S., {Sullivan}, M., {et~al.} 2002, \apj, 577, 120, (The
  Supernova Cosmology Project)

\bibitem[{Pain {et~al.}(1996)Pain, Hook, Deustua, Gabi, Goldhaber, Groom, Kim,
  Kim, Lee, Pennypacker, Perlmutter, Small, Goobar, Ellis, McMahon, Glazebrook,
  Boyle, Bunclark, Carter, \& Irwin}]{pain1996}
Pain, R., Hook, I.~M., Deustua, S., {et~al.} 1996, \apj, 473, 356, (The
  Supernova Cosmology Project)

\bibitem[{{Palanque-Delabrouille} {et~al.}(1998){Palanque-Delabrouille},
  {Afonso}, {Albert}, {Andersen}, {Ansari}, {Aubourg}, {Bareyre}, {Bauer},
  {Beaulieu}, {Bouquet}, {Char}, {Charlot}, {Couchot}, {Coutures}, {Derue},
  {Ferlet}, {Glicenstein}, {Goldman}, {Gould}, {Graff}, {Gros}, {Haissinski},
  {Hamilton}, {Hardin}, {de Kat}, {Lesquoy}, {Loup}, {Magneville}, {Mansoux},
  {Marquette}, {Maurice}, {Milsztajn}, {Moniez}, {Perdereau}, {Prevot},
  {Renault}, {Rich}, {Spiro}, {Vidal-Madjar}, {Vigroux}, {Zylberajch}, \& {The
  EROS Collaboration}}]{palanque-delabrouille1998}
{Palanque-Delabrouille}, N., {Afonso}, C., {Albert}, J.~N., {et~al.} 1998,
  \aap, 332, 1

\bibitem[{Phillips(1993)}]{phillips1993}
Phillips, M.~M. 1993, \apjl, 413, 105

\bibitem[{{Poggianti}(1997)}]{poggianti1997}
{Poggianti}, B.~M. 1997, \aaps, 122, 399

\bibitem[{{Regnault} {et~al.}(2001){Regnault}, {Aldering}, {Blanc}, {Conley},
  {Dahlen}, {Deustua}, {Ellis}, {Fan}, {Folatelli}, {Frye}, {Garavini},
  {Gates}, {Goldhaber}, {Goldman}, {Goobar}, {Groom}, {Hardin}, {Hook}, {Kent},
  {Kim}, {Kim}, {Knop}, {Lidman}, {Mendez}, {Miller}, {Moniez}, {Mourao},
  {Newberg}, {Nobili}, {Nugent}, {Pain}, {Perdereau}, {Perlmutter}, {Quimby},
  {Rich}, {Richards}, {Ruiz-Lapuente}, {Schaefer}, {Walton}, \& {Supernova
  Cosmology Project Collaboration}}]{regnault2001}
{Regnault}, N., {Aldering}, G., {Blanc}, G., {et~al.} 2001, American
  Astronomical Society Meeting, 199

\bibitem[{Riess {et~al.}(1999)Riess, Kirshner, Schmidt, Jha, Challis,
  Garnavich, Esin, Carpenter, Grashius, Schild, Berlind, Huchra, Prosser,
  Falco, Benson, Briceño, Brown, Caldwell, dell'Antonio, Filippenko, Goodman,
  Grogin, Groner, Hughes, Green, Jansen, Kleyna, Luu, Macri, McLeod, McLeod,
  McNamara, McLean, Milone, Mohr, Moraru, Peng, Peters, Prestwich, Stanek,
  Szentgyorgyi, \& Zhao}]{riess1999}
Riess, A.~G., Kirshner, R.~P., Schmidt, B.~P., {et~al.} 1999, \aj, 117, 707

\bibitem[{{Ruiz-Lapuente} \& {Canal}(1998)}]{ruiz-lapuente1998}
{Ruiz-Lapuente}, P. \& {Canal}, R. 1998, \apjl, 497, 57

\bibitem[{{Sadat} {et~al.}(1998){Sadat}, {Blanchard}, {Guiderdoni}, \&
  {Silk}}]{sadat1998}
{Sadat}, R., {Blanchard}, A., {Guiderdoni}, B., \& {Silk}, J. 1998, \aap, 331,
  L69

\bibitem[{Schlegel {et~al.}(1998)Schlegel, Finkbeiner, \& Davis}]{schlegel1998}
Schlegel, D.~J., Finkbeiner, D.~P., \& Davis, M. 1998, \apj, 500, 525

\bibitem[{Shectman {et~al.}(1996)Shectman, Landy, Oemler, Tucker, Lin,
  Kirshner, \& Schechter}]{shectman1996}
Shectman, S.~A., Landy, S.~D., Oemler, A., {et~al.} 1996, \apj, 470, 172

\bibitem[{Strolger {et~al.}(2002)Strolger, Smith, Suntzeff, Phillips, Aldering,
  Nugent, Knop, Perlmutter, Schommer, Ho, Hamuy, Krisciunas, Germany,
  Covarrubias, Candia, Athey, Blanc, Bonacic, Bowers, Conley, Dahlen, Freedman,
  Galaz, Gates, Goldhaber, Goobar, Groom, Hook, Marzke, Mateo, McCarthy,
  Mendez, Muena, Persson, Quimby, Roth, Ruiz-Lapuente, Seguel, Szentgyorgyi,
  von Braun, Wood-Vasey, \& York}]{strolger2002}
Strolger, L.-G., Smith, R.~C., Suntzeff, N.~B., {et~al.} 2002, \aj, 124, 2905

\bibitem[{{Tammann}(1970)}]{tammann1970}
{Tammann}, G.~A. 1970, \aap, 8, 458

\bibitem[{{Tonry} {et~al.}(2003){Tonry}, {Schmidt}, {Barris}, {Candia},
  {Challis}, {Clocchiatti}, {Coil}, {Filippenko}, {Garnavich}, {Hogan},
  {Holland}, {Jha}, {Kirshner}, {Krisciunas}, {Leibundgut}, {Li}, {Matheson},
  {Phillips}, {Riess}, {Schommer}, {Smith}, {Sollerman}, {Spyromilio},
  {Stubbs}, \& {Suntzeff}}]{tonry2003}
{Tonry}, J.~L., {Schmidt}, B.~P., {Barris}, B., {et~al.} 2003, \apj, 594, 1

\bibitem[{{Udalski} {et~al.}(2000){Udalski}, {Szymanski}, {Kubiak},
  {Pietrzynski}, {Soszynski}, {Wozniak}, \& {Zebrun}}]{udalski2000}
{Udalski}, A., {Szymanski}, M., {Kubiak}, M., {et~al.} 2000, Acta Astronomica,
  50, 307

\bibitem[{{Zucca} {et~al.}(1997){Zucca}, {Zamorani}, {Vettolani}, {Cappi},
  {Merighi}, {Mignoli}, {Stirpe}, {MacGillivray}, {Collins}, {Balkowski},
  {Cayatte}, {Maurogordato}, {Proust}, {Chincarini}, {Guzzo}, {Maccagni},
  {Scaramella}, {Blanchard}, \& {Ramella}}]{zucca1997}
{Zucca}, E., {Zamorani}, G., {Vettolani}, G., {et~al.} 1997, \aap, 326, 477

\end{thebibliography}

\end{document}